\documentclass[12pt,letterpaper]{article}
\pdfoutput=1
\usepackage{soul}
\usepackage{jheppub}
\usepackage{amsfonts}
\usepackage{amsmath}
\allowdisplaybreaks[4]         
\usepackage{amssymb} 
\usepackage{booktabs}  
\usepackage{euscript}       
\usepackage{color}      
\usepackage{subfig,tikz}
\usetikzlibrary{decorations.pathmorphing}
\usepackage{graphics}    
\usepackage[normalem]{ulem}
\usepackage{mathrsfs}
\usepackage{multirow}

\newcommand{\bea}{\begin{eqnarray}}
\newcommand{\eea}{\end{eqnarray}}
\newcommand{\ba}{\begin{eqnarray}}
\usepackage{braket}
\newcommand{\ea}{\end{eqnarray}}

\newcommand{\beq}{\begin{equation}}
\newcommand{\eeq}{\end{equation} }
\newcommand{\beqa}{\begin{eqnarray}}

\newcommand{\eeqa}{\end{eqnarray}}
\newcommand{\beqar}{\begin{eqnarray*}}
\newcommand{\eeqar}{\end{eqnarray*}}

\newcommand{\be}{\begin{equation}}
\newcommand{\ee}{\end{equation}}

\def\be{\begin{equation}}
\def\ee{\end{equation}}
\def\bea{\begin{eqnarray}}
\def\eea{\end{eqnarray}}

\title{ \boldmath Phase Transitions and Stability of Eguchi-Hanson-AdS Solitons}

\author[a]{Turkuler Durgut,}
\author[b]{Robie A.~Hennigar,}
\author[c]{Hari K. Kunduri,}
\author[d]{and Robert B.~Mann}

\emailAdd{tdurgut@mun.ca}
\emailAdd{robie.hennigar@icc.ub.edu}
\emailAdd{kundurih@mcmaster.ca}
\emailAdd{rbmann@uwaterloo.ca}

\affiliation[a]{Faculty of Science, Theoretical Physics, Memorial University of Newfoundland,
St. John’s, NL A1C 5S7, Canada \vspace{0.1cm}}
\affiliation[b]{Departament de F{\'\i}sica Qu\`antica i Astrof\'{\i}sica, Institut de
Ci\`encies del Cosmos,
 Universitat de
Barcelona,  Mart\'{\i} i Franqu\`es 1, E-08028 Barcelona, Spain \vspace{0.1cm}}
\affiliation[c]{Department of Mathematic and Statistics and Department of Physics and Astronomy, McMaster University,  Hamilton, ON, Canada, L8S 4M1 \vspace{0.1cm}}
\affiliation[d]{Department of Physics and Astronomy, University of Waterloo,  Waterloo, ON, Canada, N2L 3G1 \vspace{0.1cm}}

\abstract{The Eguchi-Hanson-AdS$_5$ family of spacetimes are a class of static,  geodesically complete asymptotically locally AdS$_5$  soliton solutions of the vacuum Einstein equations with negative cosmological constant.  They have negative mass and are parameterized by an integer $p\geq 3$ with a conformal boundary with spatial topology $L(p,1)$. We investigate mode solutions of the scalar wave equation on this background and show that, similar to AdS$_5$, the geometry admits a normal mode spectrum (i.e. solutions that neither grow or decay in time).  In addition, we also discuss other geometric properties of these soliton spacetimes, including the behaviour of causal geodesics and their thermodynamic properties.  We also point out a surprising connection with the AdS soliton. 
 }

\begin{document} 
\maketitle
\flushbottom

\section{Introduction}
\label{sec:Introduction}
A classic theorem of Wang~\cite{MR2156971} states that Anti-de Sitter (AdS) spacetime is the unique conformally compact (globally) static solution to the vacuum Einstein equations $G_{ab} + \Lambda g_{ab} =0$ where $\Lambda <0$ and with spherical conformal spatial boundary. The theorem holds in situations where the positive mass theorem for asymptotically hyperbolic manifolds is valid  (see also \cite{MR771446, Qing_2004}). This result lends support to the expectation that AdS spacetime is the appropriate `ground state' amongst the class of all solutions with the same asymptotic behaviour. A putative ground state would, in turn, be expected to be dynamically stable, and indeed studies of the wave equation on a fixed AdS background are consistent with this intuition. However, the remarkable result of Bizon et al. \cite{Bizon:2011gg} demonstrates that, in fact, AdS is nonlinearly unstable under arbitrarily small perturbations whose endpoint to the formation of black holes - that is, small amounts of energy tend to concentrate at shorter and shorter scales, rather than dissipating as in Minkowski spacetime~\cite{Bizon:2011gg, Bizon:2015pfa}.

A natural question is whether the vacuum Einstein equations with negative cosmological constant admit other static solutions that are asymptotically locally AdS but with a different conformal boundary.  The one-parameter family of AdS soliton spacetimes provides such an example with toroidal conformal spatial boundary~\cite{Galloway:2001uv, Galloway:2002ai}.  Clarkson and Mann considered the problem of finding static solutions asymptotic to a freely acting discrete quotient of AdS~\cite{Clarkson:2006zk, Clarkson:2005qx}.  They succeeded in constructing solutions in odd dimensions (referred to as Eguchi-Hanson-AdS spacetimes) that are asymptotic to AdS$_{d+1} / \mathbb{Z}_p$. These spacetimes have negative energy relative to that of pure AdS.  In five spacetime dimensions, $p \geq 3$ (see below) and spatial cross sections of the conformal boundary are lens spaces $L(p,1)$ equipped with the standard round metric.  Clarkson and Mann conjectured that these metrics are the states of lowest energy in their asymptotic class~\cite{Clarkson:2006zk}.

The Eguchi-Hanson-AdS$_5$ geometry, in addition to being static, has a local $SU(2) \times U(1)$ isometry group, which acts with three-dimensional orbits.  Hence its spatial sections belong to the biaxial Bianchi IX class of geometries. Dold exploited this symmetry to study the evolution of initial data within this symmetry class~\cite{Dold:2017hwr}.  In addition to showing that the resulting system of equations forms a well-posed initial-boundary value problem (with the fields satisfying an appropriate Dirichlet condition at conformal infinity), he rigorously proved that the maximal development of this restricted class of initial data sufficiently close to Eguchi-Hanson-AdS$_5$ data cannot form a horizon in the future.  Assuming that Eguchi-Hanson-AdS$_5$ is indeed the only static solution within its conformal class, this implies that the endpoint of the evolution must generically be a spacetime containing a naked singularity. 

In the present article, we will take a different perspective and study mode solutions to the massless Klein-Gordon equation
\begin{equation}\label{KGeq}
\Box_g \Phi =0
\end{equation} on the fixed Eguchi-Hanson-AdS$_5$ background.  One advantage of this approach is that we do not need to make any special symmetry restrictions on $\Phi$.  It easy to see that \eqref{KGeq} is separable and so it is relatively straightforward to reduce the problem to a single radial Sch\"odinger-type equation.  Since the background is static, it is straightforward to show that there is a conserved energy and hence a uniform bound for the energy associated with the field $\Phi$ in terms of its initial energy.  This  kind of bound, however, does not tell us if the field is being concentrated within a compact region as a result of some geometric mechanism (e.g. trapping). 

For simplicity we will study quasinormal mode solutions of \eqref{KGeq}.  We will show that, similar to AdS spacetime, Eguchi-Hanson-AdS$_5$ admits normal mode solutions (i.e. they neither grow nor decay in time). Our results will be based on a robust numerical approach as well as analytic methods.

In addition to the Klein-Gordon test field, we consider many other aspects of these solutions that have not been addressed to date in the literature. We begin with an analysis of the mechanical properties of these solitons. We compute their mass and show that these solutions have a non-trivial thermodynamic volume of topological origin~\cite{Kastor:2009wy, Andrews:2019hvq}. Examining the thermodynamics in the canonical ensemble, we show that there is an analog of the Hawking-Page phase transition~\cite{Hawking:1982dh}. The two relevant states in the phase transition are the Eguchi-Hanson-AdS$_5$ soliton and the black hole resulting from performing $\mathbb{Z}_p$ identifications to the spherical AdS black hole. We then study the geodesics in the spacetime, which is relevant for two reasons. Firstly, we seek to find whether or not there exists stable trapping of null geodesics (the confinement of null geodesics to a compact subregion of space) in the Eguchi-Hanson-AdS$_5$ spacetime. Stable trapping presents an obstruction to proving strong decay statements for solutions of the wave equation~\cite{Holzegel:2013kna, Benomio:2018ivy}. For example, decay might be no faster than inversely logarithmic in time, rather than inverse polynomial~\cite{Keir:2016azt}. The latter is expected to be necessary if there is any hope of demonstrating nonlinear stability. In particular, stable trapping has been shown in several examples of families of horizonless soliton spacetimes, which typically have some nontrivial spatial topology (e.g., two-cycles or `bubbles')~\cite{Keir:2016azt, Andrews:2019hvq, Gunasekaran:2020pue} or ultracompact objects~\cite{Keir:2014oka}. However, we find that stable trapping is absent in the Eguchi-Hanson-AdS$_5$ spacetime. Secondly, we investigate the light-crossing time of the geometry and find that it turns out to be relevant to understanding the spacing between overtones for the normal mode solutions of the Klein-Gordon equation.

A recurring theme throughout each aspect of our work is a connection between the Eguchi-Hanson-AdS$_5$ soliton and the AdS soliton that has not been pointed out in the literature. Namely, starting with the former with spatial boundary metric (the round $L(p, 1)$ lens space), we find that formally taking the $p \to \infty$ limit gives the AdS soliton as the limiting solution. As such, we show how the relevant quantities of the AdS soliton govern the asymptotics of the corresponding quantities for the Eguchi-Hanson soliton. 

The outline of our paper is as follows. In section~\ref{sec2}, we review the basic structure of the Eguchi-Hanson soliton  and show 
its limit is AdS soliton as $p \to \infty$. In section \ref{sec3}, we analyze the mass and thermodynamic behaviour of the soliton, and in section
\ref{sec4}, we consider the behaviour of timelike and null geodesics in this spacetime. We find that both massive and null particles oscillate between the edge of the soliton
and infinity, with no stable trapping regions. In section \ref{sec5}, we proceed with the main purpose of our paper, that of analyzing the scalar wave equation in
the Eguchi-Hanson soliton spacetime.  As an analytic solution is apparently intractable, we solve the equation numerically, checking our results against various approximations in certain limits.  Amongst our most intriguing results is that the normal modes interpolate between those of a scalar wave on the orbifold AdS$_5/\mathbb{Z}_2$ and on the AdS soliton as the parameter $p$ is varied. We close our paper with some concluding remarks in \ref{sec6}.  Several appendices contain details 
showing how we arrived at our results.
 
\section{Metric and Structure}\label{sec2}

Clarkson and Mann~\cite{Clarkson:2006zk} obtained a solution of the $D=5$ Einstein equations with negative cosmological constant, namely
\begin{equation}
R_{ab} = -\frac{4}{\ell^2} g_{ab}\;,
\end{equation} 
given by
\begin{align}\label{EHmet}
ds^2 &= - g(r) dt^2 
+ r^2 f(r) \left[d\psi + \frac{\cos(\theta)}{2} d\phi\right]^2
+ \frac{dr^2}{f(r) g(r)}
+ \frac{r^2}{4} d\Omega^2_2 \\
g(r) &= 1 + \frac{r^2}{\ell^2}
\quad \quad \text{and}  \quad  \quad 
f(r)=1-\frac{a^4}{r^4}.
\end{align} This is a cohomogeneity-one metric with local isometry group $\mathbb{R} \times SU(2) \times U(1)$. When $a=0$ this reduces to the AdS$_5$ metric with spherical boundary when  $t \in \mathbb{R}$, $r >0$, $\psi \in (0, 2\pi)$, $\theta \in (0,\pi)$ and $\phi \in (0,2\pi)$ with a standard apparent singularity at $r=0$, where the $(\psi, \theta, \phi)$ part of the metric degenerates, representing the origin of coordinates.   However, for $a \neq 0$ (we fix $a > 0$ without loss of generality)  the metric extends globally to a manifold with non-trivial topology, provided that certain regularity conditions are satisfied. The Killing vector field $\partial/ \partial \psi$ becomes degenerate at $r = a$; examining the $(r, \psi)$ sector of the geometry, absence of conical singularities requires the identification
\begin{equation}
\psi \sim \psi + \frac{2 \pi}{2\sqrt{g(a)}} \, .
\end{equation} The ensures that the geometry smoothly `pinches off' leaving a round $S^2$ of radius $a/2$.  This condition must be combined with the independent condition, arising from regularity of the constant $(t,r)$ surfaces, that demands 
\be\label{regcond}
\psi \sim \psi + \frac{2 \pi}{p} \, ,
\ee
where $p \in \mathbb{Z}$ (this ensures the geometry is that of $L(p,1)$).  Satisfying both conditions requires that 
\be \label{regularity}
a^2 = \left(\frac{p^2}{4} - 1 \right) \ell^2 \, .
\ee Thus regularity requires that we must have $p \geq 3$. This means that asymptotically the boundary metric is a lens space $L(p,1)$ ($p=1$ would be $S^3$).  Thus, we have a gravitational soliton (a geodesically complete, strictly stationary solution) that has a 2-cycle in the interior region and has a lens space as its boundary.   The above metric is often referred to as the `Eguchi-Hanson-AdS$_5$' as constant time hypersurfaces generalize the well-known four-dimensional Eguchi-Hanson gravitational instanton metric which must have $p=2$~\cite{Eguchi:1978xp}.

\subsection{The Large $p$ Limit of the Metric: AdS Soliton}
\label{largeP}

As we have just seen, the Eguchi-Hanson-AdS$_5$ soliton is characterized by a single integer $p$.  For several reasons, it will be fruitful to consider these solutions for large values of~$p$. 

Beginning with the Eguchi-Hanson-AdS$_5$ metric~\eqref{EHmet}, we  perform the following transformations
\begin{align}
t &= \frac{2 \tau}{p} \, , \quad r = a z \, , \quad  \theta = \frac{4 \rho}{p}\, , \quad \varphi = \psi - \frac{p}{2}  \phi \, ,
\end{align}
and then take the limit $p \to \infty$. The result is
\begin{equation} 
ds^2 = -z^2 d\tau^2 + \frac{\ell^2 dz^2}{z^2 f(z)} + \frac{\ell^2 z^2 f(z)}{4} d\varphi^2 + \ell^2 z^2\left[d\rho^2 + \rho^2 d\phi^2 \right] + \mathcal{O}\left(\frac{1}{p}\right)
\end{equation}
where 
\begin{equation}
f(z) = 1 - \frac{1}{z^4} \, .
\end{equation}
One can then convert the polar coordinates on the $\mathbb{R}^2$ to standard Cartesian coordinates, giving
\begin{equation}\label{adsSol} 
ds^2 = -z^2 d\tau^2 + \frac{\ell^2 dz^2}{z^2 f(z)} + \frac{\ell^2 z^2 f(z)}{4} d\varphi^2 + \ell^2 z^2\left[dx^2 + dy^2\right] + \mathcal{O}\left(\frac{1}{p}\right) \, .
\end{equation}
In the strict $p\to \infty$ limit, this is an AdS$_5$ soliton belonging to the class first reported in~\cite{Horowitz:1998ha}, in coordinates such that the location of the bubble is at $z = 1$. One can easily check that the coordinate $\varphi$ is periodic with period $2\pi$. Strictly speaking, there are several topologically distinct solitons that can be obtained from the same local metric~\eqref{adsSol} depending on identifications performed on the auxiliary flat directions $(x,y)$ --- see, e.g.,~\cite{Page:2002qc}. In this case, the soliton corresponding to the large $p$ limit of Eguchi-Hanson-AdS has no identifications on these coordinates, i.e. the spatial part of the boundary metric is $\mathbb{S}^1 \times \mathbb{R}^2$. This is the same configuration first considered in~\cite{Horowitz:1998ha}, and henceforth we will refer to this case as ``the'' AdS soliton. To the best of our knowledge, this connection between the Eguchi-Hanson-AdS$_5$ soliton and the AdS soliton has not been previously reported on\footnote{The connection between the Eguchi-Hanson and AdS soliton geometries could be inferred from the results for lensed CFT partition functions~\cite{Shaghoulian:2016gol}. We are grateful to Edgar Shaghoulian who, after this work was completed, brought this reference to our attention. }.

Regularity of the solution requires that $p \ge 3$ is an integer. Therefore, the $p \to \infty$ limit may be most cautiously considered as a `formal' limit. Nonetheless, it is difficult to overstate the utility of this result. As we will see in the subsequent sections, many of the quantities of interest cannot be evaluated exactly for the Eguchi-Hanson-AdS$_5$ soliton, but the asymptotics of these quantities can be effectively captured by the corresponding quantities for the AdS soliton. Said another way, Eguchi-Hanson-AdS$_5$ solitons for large values of $p$ behave in a manner similar to the AdS soliton.

\section{Soliton Mechanics}\label{sec3}

\subsection{Smarr Relation \& First Law}

While it is well-known that black holes satisfy a first law and Smarr relation, similar relationships can be found for solitons and soliton-black hole configurations. This was rigorously demonstrated in the asymptotically flat case in~\cite{Kunduri:2013vka}, and extended to a particular example of an asymptotically globally AdS soliton in~\cite{Andrews:2019hvq}. Here we apply these considerations to the Eguchi-Hanson-AdS$_5$ soliton. While the mass was calculated in the original manuscript~\cite{Clarkson:2006zk}, the notion of thermodynamic volume~\cite{Kastor:2009wy} --- which proves crucial for deriving the Smarr relation and first law in this case --- was at that point not developed.  We note also that considerations of extended thermodynamics have been previously carried out for Eguchi-Hanson-dS soliton in~\cite{Mbarek:2016mep}. 

Let $\xi = \partial_t$ be the stationary Killing field. It has zero divergence so $d\star \xi =0$, where $\star$ is the Hodge dual. It follows that one can write the closed four-form $\star \xi = -d \star \varpi$ for some locally defined 2-form $\varpi$, or equivalently
\begin{equation}\label{xi1}
\xi = \star d\star \varpi.
\end{equation} On the other hand a basic identity is 
\begin{equation}
d \star d \xi = 2 \star \text{Ric}(\xi) = \frac{8}{\ell^2} d \star \varpi
\end{equation} where the second equality follows from the Einstein equation $R_{ab} = -\tfrac{4}{\ell^2} g_{ab}$ and \eqref{xi1}.  This means we have the  conservation equation
\begin{equation}\label{identity1}
d \star \left[d\xi - \frac{8}{\ell^2} \varpi \right] =0
\end{equation} which we will integrate over a spatial hypersurface $t=$constant. If we introduce the basis
\begin{equation}\begin{aligned}
e^0 &= \sqrt{g} dt, \qquad e^1 = \frac{dr}{\sqrt{ fg}}, \qquad e^2 = r \sqrt{f} \left( d\psi + \frac{\cos\theta}{2} d\phi\right), \\
e^3 &= \frac{r}{2} d\theta, \qquad e^4 = \frac{r}{2} \sin\theta d\phi
\end{aligned}
\end{equation} and assume $\varpi$ takes the form
\begin{equation}
\varpi = A(r) e^0 \wedge e^1 + B(r) e^1 \wedge e^2
\end{equation} then a calculation gives
\begin{equation}
A(r) = \frac{1}{\sqrt{f}} \left( \frac{r}{4}  + \frac{C_1}{r^3} \right), \qquad B(r) = \frac{C_2}{r^2 \sqrt{g}}
\end{equation} so that
\begin{equation}
\varpi = \frac{1}{f} \left( \frac{r}{4} + \frac{C_1}{r^3} \right) dt \wedge dr + \frac{C_2}{r g} dr \wedge \left( d\psi + \frac{\cos\theta}{2} d\phi\right). 
\end{equation}
whose Hodge dual is
\begin{equation}
\star \varpi = -\frac{1}{4} \left(\frac{r^4}{4} + C_1\right) \sin\theta d\psi \wedge d\theta \wedge d\phi + \frac{C_2}{4} \sin\theta dt \wedge d\theta \wedge d\phi.
\end{equation}  Note that the degeneracy of  $\partial/ \partial \psi$ at the `centre' $r = a$ implies that
 $\star \varpi$ is not well defined there unless $C_1$ is chosen to be $C_1 = -a^4/4$.    However, as is typical --- and as we will see below --- regularity is not the correct prescription for fixing the parameter $C_1$.

Next, we integrate the closed form defined by  \eqref{identity1} over a hypersurface $\Sigma$ defined as a surface of constant time, $t = $ constant, over the region $R_0 \leq r \leq \infty$ with $R_0 > a$. On this region $\varpi$ is well-defined and we can apply Stokes' theorem. 
The identity gives
\begin{equation}\label{timesliceInt}
0 = \int_{\Sigma} d \left[\star \left(d \xi - \frac{8}{\ell^2} \varpi \right)\right]  = \int_{\partial_\infty \Sigma} \star \left(d \xi - \frac{8}{\ell^2} \varpi \right)  - \int_{\partial  \Sigma_{R_0}} \star \left(d \xi - \frac{8}{\ell^2} \varpi \right)
\end{equation} where $\partial_\infty \Sigma$ and $\partial \Sigma_{R_0} $ represent the asymptotic and inner boundaries respectively.  

Let us first focus on the contribution at conformal infinity. A calculation shows that as $r \to \infty$,
\begin{equation}
\star d\xi = \left(-\frac{r^4}{2\ell^2} + \frac{a^4}{2\ell^2} + O(1/r^2)\right) \sin\theta d\psi \wedge d\theta \wedge d\phi \, ,
\end{equation} 
and we note that the divergent term is precisely cancelled by the corresponding divergent term of $\star \varpi$ when the two terms are combined as in~\eqref{timesliceInt}. We identify a renormalized Komar mass as~\cite{Kastor:2009wy}
\be 
M_{\rm Komar} := -\frac{3}{32\pi} \int_{\partial_\infty \Sigma} \star \left(d \xi - \frac{8}{\ell^2} \varpi \right) = -\frac{3 \pi}{8 p \ell^2} \left( a^4 + 4 C_1 \right) \, .
\ee
The fact that the free parameter $C_1$ appears in the Komar mass can be understood as an ambiguity in the ground state energy. We will now fix this ambiguity.

We can ensure the integral over the asymptotic boundary evaluates to the mass by choosing $C_1$ appropriately. To calculate this we use the Ashtekar-Magnon procedure~\cite{Ashtekar:1984zz} which is well-defined in this setting. The relevant component of the Weyl tensor is 
\begin{equation}
C^t_{~r t r} = - \frac{a^4}{(r^2 + \ell^2)(r^4 - a^4)} = -\frac{a^4}{r^6} + O(r^{-8})
\end{equation} as $r \to \infty$. Setting $\Omega = \ell/r$ and defining the conformal metric $\bar{g}_{ab} = \Omega^2 g_{ab}$ with $\Omega =0$ as $r \to \infty$, the Ashtekar-Magnon mass is then defined as 
\begin{equation}
Q[\partial_t] = \frac{\ell}{16 \pi} \int_{\partial M} \bar{\mathcal{E}}^a_{~b} (\partial_t)^b dS_t 
\end{equation} 
where $dS_t$ is a constant time slice of the conformal boundary, which has a round lens space metric of radius $\ell$. The quantity $\bar{\mathcal{E}}^a_{~b}$ is the electric part of the Weyl tensor
\begin{equation}
\bar{\mathcal{E}}^a_{~b} = \frac{\ell^2}{\Omega^2} \bar{g}^{cd} \bar{g}^{ef} n_d n_f C^a_{~cbe},
\end{equation} 
with unit spacelike normal $n = d\Omega$.

Noting that $\bar{g}^{rr} = r^4 \ell^2$, it is then a straightforward matter to obtain the relevant component
\begin{equation}
\bar{\mathcal{E}}^t_{~t} = \frac{r^6}{\ell^6} C^t_{r t r} = -\frac{a^4}{\ell^6}
\end{equation} 
yielding
\begin{equation}\label{AMDmass}
M:= Q[\partial_t] = \frac{\ell}{16 \pi} \int_{\partial M} \left(-\frac{a^4}{\ell^6}\right) \ell^3 \frac{\sin \theta}{4} d\psi d\theta d\phi = -\frac{\pi a^4}{8 \ell^2 p} 
\end{equation} 
so the mass is negative, a fact already observed in~\cite{Clarkson:2006zk} --- c.~f.~eq.(10) of that work. Comparing the above with the result of the Komar integration implies $C_1 = - a^4/6$.

Once the Komar mass has been computed, the thermodynamic volume is identified by evaluating the integral of the Killing potential over the inner boundary, and taking the limit $R_0 \to a$. This gives:
\begin{equation}\label{volsol}
V = -\int_{\partial \Sigma_R} \star \varpi = \frac{\pi^2}{2p} \left(a^4 + 4 C_1 \right)  =  \frac{\pi^2 a^4}{6 p}\, . 
\end{equation}  using the choice $C_1 = -a^4/6$.  For some black hole solutions 
the thermodynamic volume can be interpreted as the volume of a Euclidean ball of radius $a$ that is removed from the spatial hypersurface \cite{Kubiznak:2016qmn}.  This interpretation is not available in this case because there is no `ball' in Euclidean space for which a lens space is its boundary. In this case, the thermodynamic volume $V$ has a topological origin: it arises not due to the presence of an horizon, but instead because the choice of constant $C_1$ that leads to the correct mass leads to a Killing potential that is not regular at the location of the bubble.

Noting that thermodynamic pressure is \cite{Kubiznak:2016qmn}
\begin{equation}\label{P}
P := -\frac{\Lambda}{8\pi} = \frac{3}{4\pi \ell^2}
\end{equation} 
then we have 
\begin{equation}
M  = -PV
\end{equation} 
which is the Smarr  formula for this system. 
Using the regularity condition \eqref{regcond} and the mass \eqref{AMDmass}  we have
\begin{equation}
dM = - \frac{\pi}{4p} \left( \frac{p^2}{4} -1 \right)^2 \ell d\ell
\end{equation}
Alternatively, using  \eqref{volsol}, \eqref{P},  and the regularity condition \eqref{regcond}, we have
\begin{equation}
V dP = - \frac{\pi}{4p} \left( \frac{p^2}{4} -1 \right)^2 \ell d\ell
\end{equation}
and consequently
\begin{equation}
dM = V dP  
\end{equation}
which is the expression of the first law for the Eguchi-Hanson-AdS$_5$ soliton.

The fact that the regularity condition is required for the validity of the first law is consistent with previous studies of (extended) mechanics of smooth geometries~\cite{Kunduri:2013vka, Gunasekaran:2016nep, Andrews:2019hvq}. It is worth remarking that in some cases, e.g., for accelerating black holes or spacetimes containing Misner strings, it is possible to formulate the Smarr relation and first law without requiring the regularity condition to hold~\cite{Appels:2016uha, Bordo:2019tyh}. It may be interesting to better understand when, exactly, regularity of the geometry is crucial for formulating a sensible first law and Smarr relation. 

\subsection{Euclidean action} 

By sending $t \to i \tau$  the Eguchi-Hanson-AdS$_5$ solution~\eqref{EHmet} may be analytically continued to produce a Riemannian (positive signature) Einstein metric:
\begin{equation}
ds^2 = g(r) d\tau^2 
+ r^2 f(r) \left[d\psi + \frac{\cos \theta}{2} d\phi\right]^2
+ \frac{dr^2}{f(r) g(r)}
+ \frac{r^2}{4} d\Omega^2_2 .
\end{equation} The geometry is smooth and complete with an $S^2$-bolt at $r = a$ provided the regularity condition~\eqref{regularity} is imposed.  We now periodically identify the $\tau$-coordinate  as $\tau \sim \tau + \beta$ so that it parameterizes an $S^1$.  The vector field $\partial_\tau$ is nowhere vanishing since $g(r) > 0$ and therefore this $S^1$ does not degenerate. In particular there is no condition on $\beta$.  The underlying manifold will therefore be $S^1 \times T^* S^2$ (the latter factor being the cotangent bundle of $S^2$). The metric is conformally compact with conformal boundary $S^1 \times L(p,1)$ equipped with the conformal boundary metric
\begin{equation}
\gamma = d\tau^2 +  \ell^2 \left[ \left( d\psi + \frac{\cos\theta}{2} d\phi \right)^2 + \frac{d\Omega_2^2}{4} \right].
\end{equation} The metric on the boundary $L(p,1)$ is the round metric.  

We may easily produce a Riemannian Einstein metric with the same conformal boundary by taking  appropriate angular identifications of the Euclidean Schwarzschild-AdS$_5$ metric to obtain an Einstein metric on $\mathbb{R}^2 \times L(p,1)$: 
\begin{align}\label{ESBH}
ds^2 &= U(r) d \tau^2 + U(r)^{-1} d r^2 + r^2 \left[ \left( d\psi + \frac{\cos\theta}{2} d\phi \right)^2 + \frac{d\Omega_2^2}{4} \right]
\end{align} 
where $U(r) = 1 - \mu/r^2 + r^2/\ell^2 $.  We take, as above, $\psi \sim \psi + 2\pi / p$ and $\theta \in (0, \pi)$, $\phi \in (0, 2\pi)$  and $r > r_+$ where $r_+$ is the largest root of $U(r)$. As is well known, regularity at $r=r_+$, the largest root of $U(r)$, requires that the angle $\tau$ must be identified as $\tau \sim \tau + \beta$ with 
\begin{equation}
\beta = \frac{2 \pi \ell^2 r_+}{2 r_+^2 + \ell^2}.
\end{equation} Thus for fixed temperature $T = \beta^{-1}$ there are two possible black holes
\begin{equation}
r_+ = \frac{ \pi \ell^2 \pm \ell \sqrt{\pi^2 \ell^2 - 2 \beta^2}}{2 \beta}
\end{equation} provided $T > T_{min}$ where
\begin{equation}
T_{min} = \frac{\sqrt{2}}{\pi \ell}. 
\end{equation}  Note that rather than closing smoothly to an $S^1 \times S^2$ `bolt' as in Euclidean Eguchi-Hanson-AdS$_5$, the above space has a $L(p,1)$ bolt. 

We now follow the standard procedure~\cite{Emparan:1999pm, Balasubramanian:1999re} to compare the finite Euclidean on-shell actions for these two possible infilling metrics for fixed temperature $T= \beta^{-1}$.  The renormalized Euclidean action is
\begin{equation}
I = -\frac{1}{16\pi G} \left[ \int_M \left(R_g + \frac{12}{\ell^2}\right) d \text{Vol}(g) + 2\int_{\partial M} \left( \text{Tr} K - \frac{3}{\ell} - \frac{\ell R_h}{4}\right)  d \text{Vol}(h)\right]
\end{equation} where $h$ is the metric induced on a hypersurface  $r = R$ and $R_g, R_h$ are the respective scalar curvatures of the metrics $g$ and $h$. 

 For Eguchi-Hanson-AdS$_5$ we have
\begin{equation}
I_{EH} = \frac{\beta \ell^2 \text{Vol}(L(p,1))}{16 \pi G} \left[ \frac{3}{4} - \left(\frac{p^2}{4} - 1\right)^2 \right]
\end{equation} where $\text{Vol}(L(p,1)) = 2 \pi^2/p$ is the volume of the boundary $L(p,1)$.   For the Euclidean black hole metric \eqref{ESBH} a computation gives
\begin{equation}
I_{BH} = \frac{\beta \ell^2 \text{Vol}(L(p,1))}{16 \pi G} \left[\frac{3}{4}  + \frac{r_+^2}{\ell^2} \left(1 - \frac{r_+^2}{\ell^2} \right) \right]. 
\end{equation}
With the actions at hand, simple calculations reveal that the situation is analogous to the Hawking-Page transition~\cite{Hawking:1982dh}. At low temperature the Eguchi-Hanson-AdS$_5$ soliton has the least action and dominates the canonical ensemble, whereas at sufficiently large temperatures, it is a large black hole that dominates. When the actions are equal there is a transition analogous to the Hawking-Page transition. The temperature at which the phase transition occurs can be shown to be
\be 
T_{\rm EHB} = \frac{1}{2 \pi \ell} \frac{4 + \sqrt{p^4 - 8 p^2 + 20}}{\sqrt{2 + \sqrt{p^4 - 8p^2 + 20}}} \, .
\ee
At large values of $p$ this has the asymptotic form
\be 
T_{\rm EHB}  = \frac{p}{2 \pi \ell} \left[1 + \frac{1}{p^2}  + \frac{5}{2p^4} +\cdots  \right] \, .
\ee
In the strict $p \to \infty$ limit, this phase transition is related to that which occurs between the toroidal AdS black hole and the AdS soliton~\cite{Surya:2001vj}. 

\section{Geodesics}
\label{sec4}

In this section we consider the behaviour of null and timelike geodesics in the Eguchi-Hanson-AdS$_5$ geometry.  Our primary interest is to investigate instabilities in this horizonless spacetime.  As is now well established, there is a close connection between the geometrically induced stable trapping of null geodesics and instabilities due to the clumping of wave energy.  In the geometric optics approximation, the propagation of solutions to the wave equation in a fixed background can be described by the trajectory of null geodesics.  Stable trapping is a phenomena by which properties of the geometry create obstacles forcing null geodesics to be confined in a spatially compact region.  

The prototypical example of null trapping is the photon sphere at $r = 3M$ in the Schwarzschild spacetime; null geodesics exist in circular orbits at fixed radius. Such trapping, however is unstable because any small perturbation of the orbit will cause the null geodesics to either fall towards the horizon or escape to infinity. By  contrast, stable trapping occurs when small perturbations of the orbits remain small so the trapping region is `attractive'.  Stable trapping has been shown to lead to inverse-logarithmic time decay of wave energy in both asymptotically flat~\cite{Benomio:2018ivy, Keir:2016azt, Gunasekaran:2020pue} and asymptotically AdS spacetimes~\cite{Holzegel:2013kna}. As discussed in the introduction,  this is suggestive of a non-linear instability, as linear stability typically requires decay that is an inverse polynomial function of time.

 To exploit the local $\mathbb{R} \times SU(2) \times U(1)$ isometry of the Eguchi-Hanson-AdS$_5$ spacetime, it is convenient to \eqref{EHmet} in the form
\begin{align}
ds^2 &= - g(r) dt^2 
+ \frac{dr^2}{f(r) g(r)}	
+ \frac{r^2 f(r)}{4} [d\bar\psi + \cos(\theta) d\phi]^2
+ \frac{r^2}{4} d\Omega^2_2 
\end{align}
where we have defined a new coordinate $\bar{\psi} = 2 \psi$. Then $(\bar{\psi}, \theta, \phi)$ become the familiar Euler angles, with $\theta \in (0,\pi), \bar{\psi} \in (0,4\pi/p), \phi \in (0,2\pi)$, and the metric can be expressed as 
\begin{equation}
ds^2 = - g(r) dt^2 
+ \frac{dr^2}{f(r) g(r)}	
+ \frac{r^2}{4} \left(\sigma_1^2 + \sigma_2^2 \right) 
+ \frac{r^2 f(r)}{4} \sigma_3^2 	
\end{equation}
where $\sigma_i$ are left-invariant one-forms on $SU(2)$ defined by
\begin{align}
\sigma_1 =&  - \sin \bar\psi d\theta + \cos \bar\psi \sin \theta d\phi \, ,  \\
\sigma_2 =&  \cos \bar\psi d\theta + \sin \bar\psi \sin \theta d\phi  \, ,  \\
\sigma_3 =&  d\bar\psi + \cos \theta d\phi  \, .
\end{align} Explicitly, the spatial Killing vector fields are given by
\begin{align}
R_1 &= \cot\theta \cos \phi \partial_\phi + \sin \phi \partial_\theta - \frac{\cos\phi}{\sin\theta} \partial_\psi \\ R_2 &= -\cot\theta \sin \phi \partial_\phi + \cos\phi \partial_\theta + \frac{\sin\phi}{\sin\theta} \partial_\psi \\
R_3 & = \partial_\phi, \qquad L_3 = \partial_\psi.
\end{align} The trajectories of geodesics of mass $M$ are easily found using this symmetry and the Hamilton-Jacobi method as outlined in \cite{Kunduri:2005zg}.  In particular the Hamiltonian for the motion of uncharged particles is $H = g^{ab} p_a p_b$ where $p_a$ are the canonical momenta.  The Hamiltonian system is Liouville integrable as there are five Poisson commuting functions associated with the local isometries  (there is an additional conserved quantity associated with a reducible Killing tensor). The Hamilton-Jacobi equation is 
\begin{equation}
\frac{\partial S}{\partial \lambda} + g^{ab} \frac{\partial S}{\partial x^a} \frac{\partial S}{\partial x^b} =0
\end{equation} where $\lambda$ is an affine curve parameter and it is clear one can express   the Hamilton-Jacobi function $S$ in the separable form
\begin{equation}
S = M^2 \lambda  - E t + p_\psi \psi + p_\phi \phi + \Theta(\theta) + R(r)
\end{equation}
 where  $(E, p_\psi, p_\phi)$ correspond to conserved energy and angular momenta along particle trajectories, with $p_a = \partial_a S$. Omitting details, we simply present the resulting curve equations for $x^a(\lambda)$: 
\begin{align}
\dot{t} &= \frac{2 E}{g(r)}  \qquad  \dot{\bar{\psi}} = - \frac{8 \cot \theta }{r^2} \left[\frac{p_\phi}{\sin \theta } - \cot \theta p_{\bar{\psi}} \right]
\nonumber\\
\dot{\phi} &= \frac{8}{r^2 \sin \theta} \left[\frac{p_\phi}{\sin \theta} - \cot \theta p_{\bar{\psi}} \right]  
\end{align}
and 
\begin{align}\label{geodesic}
\dot{r}^2 &= 4 E^2 f(r) - \frac{16 g(r)}{r^2} p_{\bar{\psi}}^2 - 4 M^2 f(r) g(r) - \frac{16 C}{r^2} f(r) g(r)  
\\
\dot{\theta}^2 &= \frac{64}{r^2} \left[C - \left(\cot \theta p_{\bar{\psi}} - \frac{p_\phi}{\sin \theta} \right)^2 \right]  
\end{align}
where $C$ is another constant of the motion associated with the existence of the reducible Killing tensor.  

We will now perform more detailed studies of the geodesics. 

\subsection{Time-like Geodesics: Negative Mass Repulsion}

Let us begin with a consideration of time-like geodesics, i.e.~$M \neq 0$. To illustrate some similarities/differences with time-like geodesics in AdS, we will restrict attention here to radial time-like geodesics. After defining $\mathcal{E} = E/M$ and rescaling the affine parameter accordingly, we arrive at the equation
\be 
\dot{r}^2 = f(r) \left[\mathcal{E}^2 - g(r) \right] \, .
\ee
We immediately see that the large$-r$ turning point of the motion is exactly the same as it is for AdS\footnote{Of course, the radial coordinate for the soliton is not the same as the radial coordinate for pure AdS. These differences disappear at sufficiently large $r$, as can be confirmed by putting the metric into Fefferman-Graham form. What we mean here is that the functional form of $r_{\rm max}$ is identical.}:
\be\label{maxR} 
r_{\rm max} = \ell \sqrt{\mathcal{E}^2 - 1} \, .
\ee
This result is sensible --- the space is asymptotically locally AdS and so at large enough distances the radial motion should approach that of AdS. There is a further constraint to consider since the only physically relevant cases are those for which $r_{\rm max} \ge a$. This in turn enforces that the energy must be larger than a given threshold,
\be 
\mathcal{E} \ge \mathcal{E}_{\rm min} = \frac{p}{2} \, , \quad \text{where} \quad \frac{a}{\ell} = \sqrt{\frac{p^2}{4} - 1} \, . 
\ee
A second turning point arises due to the presence of the bubble, this is at $r = a$, where $f(r) = 0$. The motion of a massive particle is therefore oscillatory, bouncing back and forth on the interval $a \le r \le r_{\rm max}$.

The presence of the bubble has implications for the motion at smaller values of $r$. To highlight this,  consider the acceleration
\be 
\ddot{r} = - \frac{r}{\ell^2} - \frac{a^4}{r^3 \ell^2} + \frac{2 a^4(\mathcal{E}^2 - 1)}{r^5}  
\ee
where the first term on the right-hand side is the acceleration term present in pure AdS. This makes manifest the well-known fact that the motion of time-like geodesics in AdS is periodic with period $2 \pi \ell$. Further, note that the sign of the acceleration in pure AdS is \textit{always negative} --- a stone tossed in AdS will always be returned to sender. 

The additional acceleration terms for $a \neq 0$ have interesting consequences. The last term in the above always makes a positive contribution, since $\mathcal{E} \ge \mathcal{E}_{\rm min} > 1$. Close to the bubble this positive term actually dominates,  leading to a region where the acceleration is positive, indicating a repulsion. 

It is a simple matter to prove this. First, note that sufficiently large $r$, the leading AdS term will always dominate, meaning that $\ddot{r} < 0$ at large $r$. Next, let us consider the possibility of solutions to the equation $\ddot{r} = 0$.  Define a new variable $r = \ell(x + \alpha)$ where $\alpha = a/\ell$. Then, the equation $\ddot{r} = 0$ becomes equivalent to the polynomial equation\footnote{While an explicit solution for $x$ in this equation can be obtained, it is sufficiently complicatedly that it is not beneficial to present it here. }
\be 
x^6 + 6 \alpha x^5 + 15 \alpha^2 x^4 + 20 \alpha^3 x^3 + 16 \alpha^4 x^2 + 8 \alpha^5 x + 2 \alpha^4 \left(1-\mathcal{E}^2 + \alpha^2\right) = 0 \, .
\ee
Every term in this polynomial is manifestly positive except for the very last one. It will vanish for
\be 
\mathcal{E} = \mathcal{E}_{\rm min} = \frac{p}{2}  \, .
\ee
For $\mathcal{E} > \mathcal{E}_{\rm min}$ the last term is negative. In this case, applying Descartes' rule of signs tells us that there will be a single positive value of $x$ where the above polynomial has a zero. There are no zeros for positive $x$ under other circumstances. Undoing our substitutions, $x > 0$ implies $r > a$. Thus, we have concluded that there is exactly one zero for the acceleration for $r > a$. Since we know from the above analysis that $\ddot{r} < 0$ for sufficiently large $r$, we then conclude that in a neighbourhood of the bubble the acceleration is positive for  particles with $\mathcal{E} > p/2$. Since this corresponds to the minimum possible energy, it follows that \textit{all} massive particles feel a repulsion in the vicinity of the bubble, except those for which $\mathcal{E}$ is \textit{exactly} $\mathcal{E}_{\rm min}$ as these particles just sit at the bubble without motion.

The extent of this repulsion is bounded and approaches a constant, as can be seen by expanding the acceleration in the vicinity of the bubble:
\be 
\ddot{r} = \frac{(4\mathcal{E}^2-p^2)}{\ell \sqrt{p^2-4}} + \mathcal{O}(r-a) \, .
\ee
The origin of this effect is the negative mass of the bubble. To see this, it is instructive to re-write the expression for the acceleration in terms of the mass given in~\eqref{AMDmass}. It is a simple matter to show that it then takes the form,
\be 
\ddot{r} = - \frac{r}{\ell^2} - \frac{8 p M }{ \pi} \frac{\left( 2 r_{\rm max}^2 - r^2 \right)}{r^5} \, ,
\ee
where $r_{\rm max}$ was introduced in eq.~\eqref{maxR}. Since $r \le r_{\rm max}$ we see directly that if it were possible to have positive mass, then the acceleration would always be attractive. However, since the mass is necessarily negative, there is a competition between the confining potential of AdS and the negative mass repulsion of the bubble. This leads to a thin layer in the vicinity of the bubble where massive particles find themselves accelerated away from the bubble.

Finally, it is important to emphasize that the repulsion does not result in a situation where a (positive energy) particle `hovers' at some fixed position $r > a$. When $\mathcal{E} > \mathcal{E}_{\rm min}$ the velocity is necessarily non-zero at the point where the acceleration vanishes. The only case when it is possible for $\dot{r} = \ddot{r} = 0$ simultaneously is when $\mathcal{E} = \mathcal{E}_{\rm min}$, corresponding to a particle at the location of the bubble.

The (radial) motion of massive particles in Eguchi-Hanson-AdS$_5$ is therefore qualitatively similar to the motion of massive particles in AdS. The motion is forever oscillatory, with particles confined in some layer surrounding the bubble, the thickness of which depends on the energy of the particle. Unsurprisingly, this is suggestive that Eguchi-Hanson-AdS$_5$ may suffer from a similar non-linear instability as global AdS, however, here without the possibility of forming horizons~\cite{Dold:2017hwr}.

\subsection{Null Geodesics: Absence of Stable Trapping}
Consider, now,  null geodesics.   These are obtained by setting $M=0$ in \eqref{geodesic}, yielding
\begin{equation}
\dot r^2 = 4  f(r) - \frac{16 g(r) \eta^2}{r^2} - \frac{16 \hat C}{r^2} f(r) g(r)
\end{equation} where $\eta:= p_\psi/E$ and $\hat C = C/E^2$ and we have performed an appropriate rescaling of the affine parameter.   For convenience, we will work with dimensionless parameters by scaling out $\ell$ (this is equivalent to fixing $\ell = 1$). Doing so produces  
\begin{equation}
\dot r^2 = V(x) = \frac{4}{x^3}P(x), \qquad P(x) = c_3 x^3 + c_2 x^2 + c_1 x + c_0 .
\end{equation} where $x = r^2$ and
\begin{equation}
c_3 = 1 - 4 \hat{C} - 4 \eta^2, \quad c_2 = -4(\hat{C} + \eta^2), \quad c_1 = -a^4(1 - 4\hat{C}), \quad c_0  = 4 a^4 \hat{C}.
\end{equation} Trajectories are only allowed in regions where the effective potential $V(r) > 0$ with turning points at the zeroes, whereas regions with $V(r) < 0$ are forbidden. Stable trapping will occur if there exist $x_1, x_2$ such that $0 < a^2 \leq x_1 < x_2$ with $V(x_i) = 0$, $V(x) > 0$ for $x \in (x_1, x_2)$, and $V(x) <0$ in  neighbourhoods to the left  of $x_1$ and right of $x_2$ (see~\cite[Fig 2]{Gunasekaran:2020pue}).   This translates into similar conditions on the cubic $P(x)$. First note that $\hat{C} \geq 0$ by definition. Observe that 
\begin{equation}
P(0) = 4 a^4 \hat{C} \geq 0, \qquad P(a^2) = -4 a^4 ( 1 + a^2) \eta^2 \leq 0 .
\end{equation} We now consider several distinct cases. 
\subsubsection{Case 1: $\eta, \hat{C} \neq 0$} From above, we have $P(0) > 0$ and $P(a^2) <0$. Thus there must be at least 1 root $x_0$ with $0 < x_0 < a^2$.  For stable trapping we will need two further positive roots $x_1, x_2$ each strictly greater than $a^2$. As $x \to \infty$, the sign of $P$ is controlled by $c_3$.  If $c_3 > 0$ then it is clear there cannot be a 2nd root $x_2$.  This occurs if
\begin{equation}
0 < \hat{C} < \frac{1}{4} - \eta^2 < \frac{1}{4} 
\end{equation} Thus we find there is no stable trapping in this case. Now, suppose that $c_3 < 0$. then 
\begin{equation}
\hat{C} > \frac{1}{4} - \eta^2
\end{equation}  Now consider Descartes' rule of signs. We are assuming $c_3 < 0$, and $c_2 < 0$, whereas $c_0 >0$ automatically. Thus there is only one sign flip between adjacent coefficients:  if $c_1 > 0$ there is a sign flip between the $x^2$ and $x$ coefficients, and if $c_1 < 0$ there is a  sign flip between the $x$ and $x^0$ coefficients. Thus the rule of signs indicates there can be only one positive root, but we already know one exists in $(0, a^2)$. Thus stable trapping cannot occur in this case either. 

Finally suppose $c_3 =0$ so that
\begin{equation}
\hat{C} = \frac{1}{4} -\eta^2 > 0 
\end{equation} 
in which case we must have $\eta^2 < 1/4$.  Writing $\eta^2 = 1/4 - \epsilon$ for $0 < \epsilon < 1/4$, we have 
\begin{equation}
P(x) = -x^2 + a^4 (4\epsilon -1) x + 4 a^4 \epsilon
\end{equation} 
which is a downward-pointing parabola. 
 It is easy to see that the only turning point has to be for $ x < 0$;  in particular there cannot be two roots to the right of $x = a^2$. Thus there is no stable trapping here either. 
\subsubsection{Case 2: $\hat{C}=0$} In this case we get
\begin{equation}
P(x) = x ((1 - 4\eta^2)x^2 - 4\eta^2x - a^4).
\end{equation} Then $x=0$ is automatically a root. Suppose that $\eta^2 =0$. Then $P = x(x-a^2)(x+a^2)$ for which we easily read off the roots and find stable trapping cannot occur. Hence assume $\eta^2 > 0$.   To get stable trapping we need the quadratic $Q(x) = (1 - 4\eta^2)x^2 - 4\eta^2 x - a^4$ to have two roots that are greater than $a^2$.  From Descartes' rule of signs we see that there can be at most 1 sign flip between adjacent coefficients, and hence only one positive root. Thus there is no stable trapping here either. 
\subsubsection{Case 3: $\eta =0$} We have
\begin{equation}
P(x) = (1 - 4\hat{C}) (x- a^2)(x+a^2) \left(x - \frac{4\hat{C}}{1 -4\hat{C}} \right)
\end{equation}  If $C = 1/4$ then this is just
\begin{equation}
P = -(x - a^2)(x+a^2)
\end{equation} which is a downward pointing parabola and there is no stable trapping here. Hence assume $\hat{C} \neq 1/4$.  There is a root $x_- = -a^2$ and $x_1 = a^2$.  Then the only way we have the final root $x_2$ to the right of $x = a^2$ is if 
\begin{equation}
\frac{4\hat C}{1 - 4 \hat C}  > a^2
\end{equation} which, since $\hat C>0$ implies we must have $1 - 4\hat C > 0$. But then $P'(x_2) > 0$, which cannot occur for stable trapping. 

The above cases exhaust all possibilities, establishing that stable trapping does not occur. 

\subsection{Null Geodesics: Light-Crossing Time}

In AdS, a light ray sent from a given point completes a round-trip to infinity and back in finite time coordinate time. Taking this point to be the orign, we have
\be 
T_{\rm AdS} = 2 \int_{r=0}^{r=\infty} \frac{\dot{t}}{\dot{r}} = 2 \int_0^\infty \frac{1}{g(r)} = \pi \ell \, .
\ee
This in turn defines a fundamental frequency naturally associated with AdS:
\be 
\omega_{\rm AdS}\ell = \frac{2 \pi \ell}{T_{\rm AdS}} = 2 \, .
\ee
This is relevant because the fundamental frequency matches the spacing for the overtones of scalar normal modes in AdS. Since we will study scalar normal modes in the next section, it is relevant to understand if a similar effect occurs for the soliton. 

The computation for the light-crossing time in the soliton proceeds along the same lines. The integral to evaluate is now
\be\label{lightCross} 
T_{\rm p} = 2\int_a^\infty \frac{1}{g(r) \sqrt{f(r)}} dr \, ,
\ee
where here the subscript $p$ refers to the integer defining regularity of the geometry. This integral actually has a closed form expression,
\begin{align}\label{lc_eval}
\frac{T_p}{\ell} &= \frac{1}{4 8  \sqrt{2 \pi} (p^2-4)^{3/2}} \left\{48(p^2-4)^2 \Gamma\left(\frac{3}{4} \right)^2 \left[ {}_2 F_1 \left(-\frac{1}{4},1,\frac{1}{4}, \frac{16}{(p^2-4)^2}  \right) -1 \right] \right.
\nonumber\\
&\left. \, +  \Gamma\left(\frac{1}{4} \right)^2 \left[48 (p^2-4) + \frac{256}{p \sqrt{p^2-8}} {}_2F_1 \left(\frac{1}{2},\frac{3}{4},\frac{7}{4}, \frac{16}{(p^2-4)^2}  \right) \right]\right\} \, .
\end{align}
Expanding this in different limits is more useful. First, for $p \to 2$, the expansion is
\be 
\frac{T_p}{\ell} = \pi - \sqrt{\frac{2}{\pi}} \Gamma\left[\frac{3}{4} \right]^2 \sqrt{p-2} - \frac{4\pi \Gamma\left[1/4 \right] + 3 \sqrt{2} \Gamma[3/4]^3}{24 \sqrt{\pi} \Gamma[3/4]} (p-2)^{3/2} + \frac{\pi}{2} (p-2)^2 + \mathcal{O}(p-2)^{5/2} \, .
\ee
The other limit of interest where this can be expanded is $p \to \infty$. In this case the expansion reads,
\be 
\frac{T_p}{\ell} = \frac{\Gamma[1/4]^2}{\sqrt{2 \pi} p} + \sqrt{\frac{2}{\pi}} \frac{\Gamma[1/4]^2 - 8 \Gamma[3/4]^2}{p^3} + \frac{3 \left(17 \Gamma[-3/4]^2 - 256 \Gamma[3/4]^2 \right)}{8 \sqrt{2 \pi} p^5} + \mathcal{O}(p^{-6}) \, .
\ee

\begin{figure}
\centering
\includegraphics[width=0.66\textwidth]{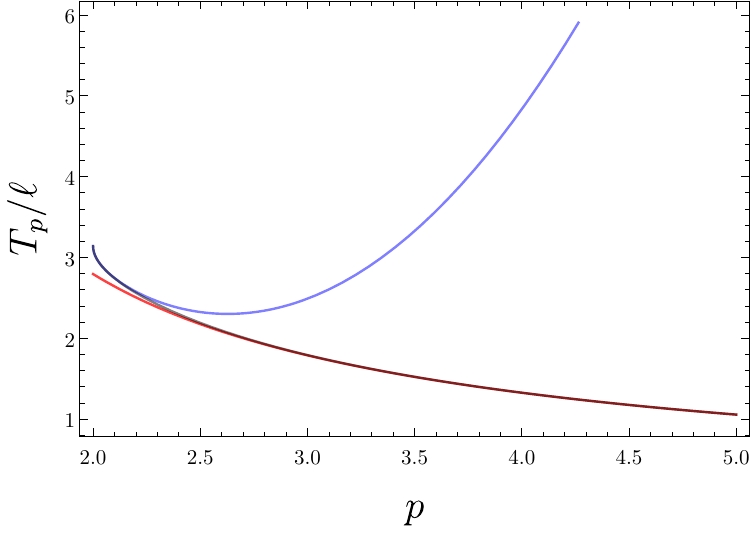}
\caption{A plot of the light-crossing time $T_p$ as a function of $p$ (black curve) compared with the asymptotic approximations as $p \to 2$ (blue curve) and as $p \to \infty$ (red curve). In each case we include the first three terms of the approximation.  We see that the large $p$ approximation   is quite accurate for all physical values of $p \ge 3$.}
\label{light-cross}
\end{figure}

We show in Figure~\ref{light-cross} a plot of the exact evaluation of the light crossing time, compared with the approximate forms described above. Here we have plotted the result treating $p$ as a continuous parameter. It must be kept in mind that regular geometries exist only for integer $p \ge 3$. First, we note that the light-crossing time for the soliton is always less than that of pure AdS. Regarding the approximations, for $p \sim 2$ the approximation is accurate in the close vicinity of $p = 2$, but is a poor approximation for larger, physical values of $p$. The large $p$ approximation is much better suited for all cases $p \ge 3$. Finally, in the limit $p \to \infty$, we recover the light-crossing time of the AdS soliton, $\tau_{\rm AdSS} = \Gamma[1/4]^2/(2 \sqrt{2 \pi})$.\footnote{See, for example, \cite{Myers:2017sxr}.}
This is observed after performing the shift of the time coordinate necessary for that limit $\tau = p t/2$ --- see Section~\ref{largeP}.

\section{Wave Equation on Eguchi-Hanson Solitons}
\label{sec5}

A massive Klein-Gordon field $\Phi$ obeys the equation 
\begin{equation}
\nabla^\mu \nabla_\mu \Phi = M^2 \Phi 
\end{equation}
in the spacetime background metric~\eqref{EHmet}. 
We use the separation ansatz
\be 
\Phi = e^{-i \omega t} e^{i m \psi} Y(\theta, \phi) R(r) 
\ee
where $Y(\theta, \phi)$ is an eigenfunction of the charged scalar Laplacian on $CP^1$, satisfying
\begin{equation}
D^2Y(\theta, \phi) = - \mu Y(\theta, \phi) 
, .
\end{equation} 
The spectrum of this operator, and the associated eigenfunctions have been studied~\cite{Wu:1976ge,Warner:1982fs,Hoxha:2000jf}, and the spectrum is
\begin{align}
\mu=& \, l \left( l+2\right) - m^2 \, , \\
l=&  \, 2k + |m| \, , \\
k=& \, 0,1,2,\dots \, .
\end{align} 
Requiring $\Phi$  to be smooth implies  $m = n p$, as $\psi$ is identified with period $2\pi / p$. With this observation, the problem reduces to that of a single radial equation, which reads
\begin{equation}\label{radialEq}
\frac{1}{r} \frac{d}{dr} 
\left( f(r) g(r) r^3 \frac{dR(r)}{dr}\right) 
+\left[ \frac{\omega^2 r^2}{g(r)}
- \frac{ m^2}{f(r)}
- M^2 r^2 
- \mu \right] R(r) 
=0  \, .
\end{equation}

The radial equation \eqref{radialEq} can be cast into Schr\"odinger equation form. To see this, we introduce  a new independent variable $x$  
\begin{equation}
\frac{dx}{dr} = \frac{1}{g \sqrt{f}}, \qquad x(a) = 0
\end{equation} 
and a dependent variable $\Psi$  
\begin{equation}
\Psi = f^{1/4} r^{3/2} R \, .
\end{equation} 
This puts the radial equation into a formally self-adjoint form
\begin{equation}
-\frac{d^2 \Psi}{dx^2} + V(r(x)) \Psi =  \omega^2 \Psi
\end{equation} with potential
\begin{equation}
V(r) = -\frac{ g f^{1/4}}{r^{3/2}}\frac{d}{dr} \left[r^3 f g \frac{d}{dr} (f^{-1/4} r^{-3/2} ) \right] + \frac{g}{r^2} \left(\frac{m^2}{f} + \mu + M^2 r^2\right) \, .
\end{equation} 

We now seek solutions of this equation. We have not been able to find an analytic solution, and so we proceed with a combination of numerical and approximate techniques.

\subsection{Approximate Solution: WKB Analysis}

In the limit of large eigenvalues the differential equation can be solved approximately using, for example, the WKB method~\cite{bender78:AMM}. For the purposes of this analysis, it is useful to rewrite the differential equation \eqref{radialEq} by defining 
\be 
r = z a \, ,
\ee
which maps the domain to $z \in [1, \infty)$. After this transformation, the resulting equation is given by
\begin{align}\label{largep1}
0 =& z (z^4-1)\left(4 + (p^2-4)z^2\right) R''(z) + \left(4 -(p^2-4) z^2 + 12 z^4 + 5  (p^2-4) z^6 \right) R'(z) 
\nonumber\\
&+ 4 z^3 \left[ -4 k^2 - 2 n p - 4 k (1+np) + \frac{n^2 p^2 z^4}{1-z^4} + \frac{(p^2-4) z^2 \ell^2 \omega^2}{4 + (p^2 -4)z^2}\right] R(z) \, .
\end{align}

We work here in an approximation where the overtone number $N$ is larger than the other fixed quantum numbers characterizing the problem, i.e. $N \gg n, k$.  In this case, by defining a new function and variable according to
\be\label{domainStretch} 
z = 1 + e^x \, , 
\ee 
and 
\be 
\Psi = \sqrt{\frac{z}{(z+1)(z^2+1)(4 + z^2(p^2-4))}} R(z) \, ,
\ee
the differential equation takes the form of a Schr\"odinger equation,
\be 
-\frac{1}{(\omega \ell)^2} \frac{d^2 \Psi}{dx^2} + V \Psi = 0\, , 
\ee
with potential
\be 
V = \frac{4 z^4(z-1)(p^2-4)}{\left(4 + z^2(p^2-4) \right)^2 (z+1)(z^2+1)}  
\ee
 in the limit of large $(\omega \ell)$. 
The transformation \eqref{domainStretch} has mapped the problem to the interval $(-\infty,\infty)$, which allows for direct comparison with the analysis of~\cite{bender78:AMM}. Examining the potential as a function of $x$ we see that it vanishes at the boundaries $x \to \pm \infty$, and there are no turning points on the interior domain. We can then apply directly the results of~\cite[Sec. 10.5]{bender78:AMM} for the quantization condition of normal mode solutions in the geometric optics approximation:
\be 
\omega \ell \int_{-\infty}^{\infty} \sqrt{V(x)} dx = \left(N + \frac{1}{2} \right)\pi + \mathcal{O}\left(\frac{1}{\omega} \right) \, ,
\ee 
where $N$ is a non-negative integer. Written as an integral over $z$, the left-hand side of the above becomes
\be 
\int_{-\infty}^{\infty} \sqrt{V(x)} dx = \int_{1}^\infty \frac{2 z^2}{\left(4 + z^2(p^2-4)\right)} \sqrt{\frac{ p^2-4}{z^4-1}}  dz = \frac{T_p}{2 \ell} \, ,
\ee
where in the last equality we recognize the integral as being identical to that defining the light crossing time of the soliton geometry given in~\eqref{lightCross}. Therefore, applying the quantization condition as above, we have in the geometric optics approximation the following result for the frequencies:
\be 
\omega = \left(N + \frac{1}{2} \right) \frac{2\pi}{T_p} =   \left(N + \frac{1}{2} \right) \omega_{\rm fun}\, ,
\ee
where the fundamental frequency $\omega_{\rm fun} = 2 \pi /T_p$ for the Eguchi-Hanson-AdS$_5$ soliton. We expect this relationship to be accurate in the limit of large overtone number, since  we have assumed that $\ell\omega$ is large. Later, in our numerical results, we will verify this is remarkably accurate even at small overtone number.

As explained below~\eqref{lightCross}, the light crossing time can be evaluated analytically and was presented in Eq.~\eqref{lc_eval}.
While this closed-form solution for the light-crossing time (or, equivalently, the fundamental frequency) is convenient, it is not necessarily illuminating. Therefore in Table~\ref{fun_freq_vals} we list the numerical values for the fundamental frequency for a few values of $p$.  Since the asymptotics of $T_p$ are dominated by a $1/p$ term, we have factored out an overall multiple of $p$ in the numerical expressions for $\omega_{\rm fun}$. This allows one to see more clearly the limiting behaviour for larger values of $p$. The results in this table should be compared with the fundamental frequency for the AdS soliton, which is
 \be
 \ell \omega_{\rm fun}^{\rm AdSS} = \frac{4 \sqrt{2} \pi^{3/2}}{\Gamma[1/4]^2} =  \lim_{p \to \infty} \frac{2 \ell \omega_{\rm fun}}{p} \approx 2.39628  \, .
 \ee 
That is, at large values of $p$, the fundamental frequency, which governs the normal modes at large overtone number $N \gg k, n$, asymptotically approaches $p/2$ times the fundamental frequency of the AdS soliton. This is directly related to the fact that there is a sense in which the AdS soliton is formally the large $p$ limit of the Eguchi-Hanson-AdS$_5$ soliton, as explained in Section~\ref{largeP}. 
 The factor of $2/p$ is precisely the factor required to match the time coordinates between the two geometries.

\begin{table}
\begin{center}
\begin{tabular}{c c c c c c c c c c c}
 \toprule
 & & & \multicolumn{4}{l}{Numerical values of $\omega_{\rm fun}$}
 \\
 \toprule
 $p$ &  3 & 4 & 5 & 6 & 7 & 8 & 8 & 10
 \\ 
 $\ell \omega_{\rm fun}/p$ &
 1.16822
 & 1.18337
  & 1.18917
  & 1.19207
 & 1.19375 
 & 1.19481 
  & 1.19553
  & 1.19604
  \\
 \bottomrule
\end{tabular}
\end{center}
\caption{A selection of numerically obtained  $\omega_{\rm fun}$ values.}
\label{fun_freq_vals}
\end{table}

\subsection{Numerical Solution: Boundary Conditions \& Regularity}

To implement our numerical methods, it is essential to understand the behaviour of the solutions to the radial equation in the vicinity of the bubble and asymptotically. It will also be more convenient to compactify the semi-infinite domain to a finite interval. We therefore begin our numerical analysis of the radial equation with an analysis of the asymptotic behaviour of the solutions and boundary conditions. 

We begin by introducing a new coordinate  
\be\label{compactTrans} 
u = \frac{2(r-a)}{r+a} - 1 \,, 
\ee
which maps the semi-infinite domain $r \in (a, \infty)$ to the interval $u \in [-1,1]$. To understand the singularity structure, we perform a Frobenius analysis. Near $r = a$ we write,
\be 
R(u) = (u+1)^s \sum_{i= 0} a_i (u+1)^i 
\ee
and extract $s$ by solving the differential equation near $u = -1$. Provided that $m \neq 0$, the equation allows for two solutions, corresponding to the values
\be 
s = \pm \frac{m}{2 p} \, .
\ee
We require regularity of the solution as $u \to -1$, and therefore only one of the above solutions is physically acceptable, depending on the sign of $m$. In general, for $m \neq 0$ we have the regular behaviour $s = |m|/(2p)$. 

The case $m = 0$ must be treated separately, since the Frobenius method gives a degenerate root in that case. Since the degenerate root corresponds to $s = 0$, from the general theory of Frobenius analysis, we can conclude in this case that the solution must be of the form
\be 
R(u) =  \sum_{i= 0} a_i (u+1)^i + \log (u+1) \sum_{i= 0} b_i (u+1)^i 
\, .
\ee
It can easily be shown that this ansatz leads to a consistent series solution in the vicinity of $u = -1$. Regularity of the solution there forces us to set $b_0 = 0$, and so when $m = 0$, the solution approaches a constant as $r \to a$. This behaviour is in fact captured by the result above, $s = |m|/(2p)$, in the case $m=0$. 

The asymptotic analysis near $u = 1$ (i.e. $r \to \infty$) is more standard. We can again proceed via Frobenius analysis. Taking the Frobenius ansatz
\be 
R(u) = (u-1)^{\hat{s}} \sum_{i= 0} a_i (u-1)^i \, ,
\ee
expanding the differential equation near $u \to 1$ and demanding a solution of the indicial equation, we find
\be 
\hat{s} = 2 \pm \sqrt{2 + (M \ell)^2} \, .
\ee
Requiring the solution to be real gives the well-known Breitenlohner-Freedman bound~\cite{Breitenlohner:1982jf}, $(ML)^2 > -2$. As usual, we proceed by taking the normalizable solution,
\be 
\hat{s} = 2 + \sqrt{2 + (M \ell)^2} \, .
\ee

With the asymptotic behaviours understood, we now recast the differential equation into a form that is more amenable to numerical solution. To this end, we define a new function
\be\label{RtoH} 
R(u) = (1-u)^{2 + \sqrt{2 + (M \ell)^2}} (1+u)^{|m|/(2p)} h(u) \, .
\ee
We then recast~\eqref{radialEq} in terms of the new function $h(u)$. The resulting expression is somewhat messy, and so we do not present it here. The prefactors implement the appropriate fall-off conditions in the two relevant limits, and therefore the only requirement on $h(u)$ is that it should be regular. Demanding this,  a series solution near either $u \to -1$ or $u \to +1$ results in Robin boundary conditions 
\begin{align}
\label{bcs}
h'(-1) &= \left[\frac{(4 + p^2)(m^2 + 2 |m| p) + 4 p^4  +8 \omega^2 + 2 p^2 (\mu - \omega^2)}{2 p^3 (|m|+p)} \right] h(-1)  
\\
h'(1) &= \left(1 + \frac{|m|}{4 p}\right) h(1)  
\end{align}
that must be imposed on the function $h(u)$. 

Before moving on, it is worth commenting in a bit more detail about the solution in the near bubble regime $r \to a$. In this regime, the radial solution behaves as $R(r) \sim (r-a)^{n/2}$, and so it appears to be continuous but not smooth there. To study this more carefully, we carry out the transformation
\be \label{polcor}
\rho = \frac{4 \sqrt{r-a}}{p \sqrt{f'(a)}} \, , \quad \varphi = p \psi  
\ee
which yields
\be 
ds^2 = d\rho^2 + \rho^2 d\varphi^2 
\ee
namely the standard polar metric on $\mathbb{R}^2$ as $r \to a$, with $\varphi \sim \varphi + 2 \pi$ due to the $2 \pi/p$ periodicity of $\psi$. 

Consider next the solution to the wave equation as $r \to a$. Focusing just on the terms with $r$ and $\psi$ dependence, and transforming these according to the polar coordinates \eqref{polcor} defined above, we obtain
\be 
\phi(r, \psi) \sim \rho^n e^{i n \varphi} \, .
\ee
Despite the fact that neither $\rho$ nor $\varphi$ are themselves smooth functions --- as can be confirmed by transforming these quantities to a Cartesian frame --- the combinations as appearing here are indeed smooth. The solutions have the same structure as Bessel functions.

\subsection{Numerical Solution: Normal Modes}

We can now perform a numerical analysis of the radial equation~\eqref{radialEq}. To solve this equation, we have employed two different numerical schemes. Our primary method has been a pseudospectral method for obtaining the eigenvalues of the differential operator. However, we have also cross-checked and benchmarked this method with a simpler shooting method.  In Appendix~\ref{shooting} we compare the two techniques. 

For an in-depth review of the pseudospectral method we refer to various references, e.g.~\cite{Boyd, canuto2007spectral, Dias:2015nua}.  After the transformations~\eqref{compactTrans} and \eqref{RtoH}, the radial equation is a problem on the interval $[-1, 1]$. On this interval, we introduce a grid consisting of $\mathcal{N} + 1$ grid points $u_i$, which we take to be the Gauss-Lobatto points,
\be 
u_i = \cos \frac{i \pi}{\mathcal{N}}  \quad \text{for } i = 0, \dots , \mathcal{N} \, .
\ee
We then discretize the differential equation. The eigenfunction $h(u)$ becomes a vector defined at the Gauss-Lobatto points, $h_i := h(u_i)$, and the derivatives appearing in the differential equation are replaced with the corresponding Chebychev differentiation matrices\footnote{We refer to, for example, \cite{canuto2007spectral} for an explicit form of these objects --- c.f. section 2.4.2 therein.}  $D_{i,j}$. The differential equation then reduces to a generalized eigenvalue problem,
\be 
H_{i,j} h_j = \omega^2 V_{i,j} h_j 
\ee
where $H_{i,j}$ is the discretization of the differential operator and $V_{i,j}$ is the matrix discretization of the terms multiplying $\omega^2$ in the original differential equation. To implement the Robin boundary conditions, we replace the first row of the above with the discretization of the second equation in \eqref{bcs}, and the last row with the discretization of the first equation of~\eqref{bcs}.

After the differential equation has been discretized, we utilize the built-in eigenvalue solvers of Mathematica to obtain the eigenvalues $\omega$. For a given choice of $\mathcal{N}$ we will obtain a set of eigenvalues $\{\omega \}_{\mathcal{N}}$ for the generalized eigenvalue problem described above. Of course, for any discretization of the differential equation, the eigenvalues of the corresponding matrix equation will differ from the true eigenvalues of the differential operator. We expect that the error in this discretization will become smaller as $\mathcal{N}$ is increased. We monitor convergence in the following way. We set \textit{a priori} a tolerance that we regard as the minimum acceptable absolute error in the eigenvalue $\omega$. We then choose a number of grid points $\mathcal{N}$ and compute the spectrum $\{\omega\}_\mathcal{N}$ first for $\mathcal{N}$ and then the spectrum $\{\omega\}_{\mathcal{N}+1}$ for a grid with $\mathcal{N} + 1$ points. For each eigenvalue in $\{\omega\}_\mathcal{N}$ we assess whether there is a corresponding eigenvalue in $\{\omega\}_{\mathcal{N}+1}$ that is within the specified tolerance. If it is, then we conclude that this particular eigenvalue has been determined to the specified tolerance. We continue in this way, identifying each eigenvalue in $\{\omega\}_\mathcal{N}$ that has converged within the specified tolerance. For most of our results  the tolerance has been set to $10^{-5}$, meaning the eigenvalues are accurate to \textit{at least} five decimal places. In most cases, we have repeated the above procedure for several values of $\mathcal{N}$ to gain a better understanding of the convergence properties. In Appendix~\ref{p2Sol} we present additional details on convergence for particular cases. Furthermore, in Appendix~\ref{shooting} we compare the results obtained via the pseudospectral method with those obtained via the shooting method, to ensure consistency.

\begin{figure}
\centering
\includegraphics[width=0.9\linewidth]{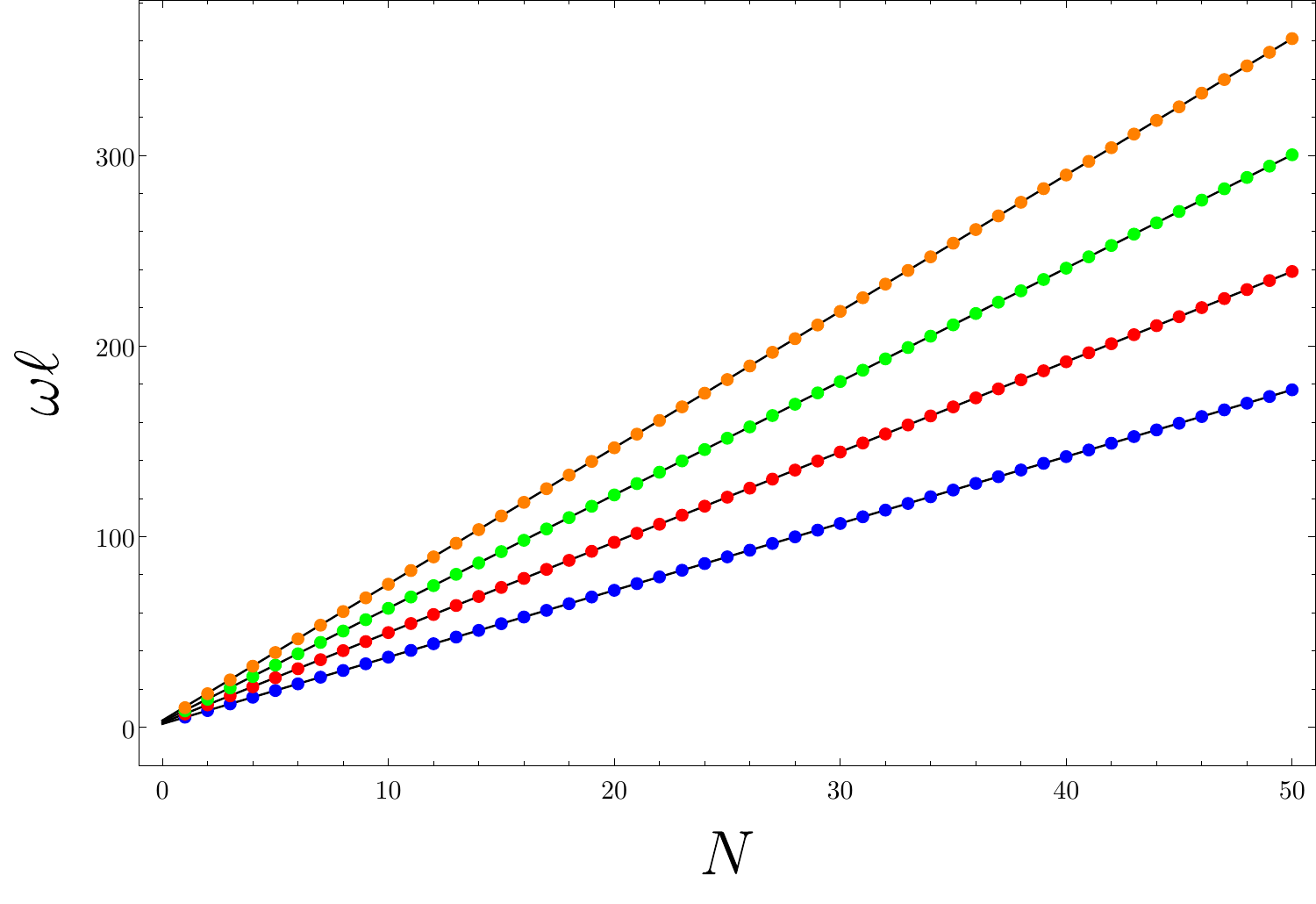}
\caption{A comparison of the WKB approximate results for $n=k=0$ (black lines) with the numerically computed eigenvalues for $p = 3,4,5,6$ corresponding to the blue, red, green and orange dots, respectively.}
\label{wkb_compare}
\end{figure}

\begin{figure}
\centering
\includegraphics[width=0.9\linewidth]{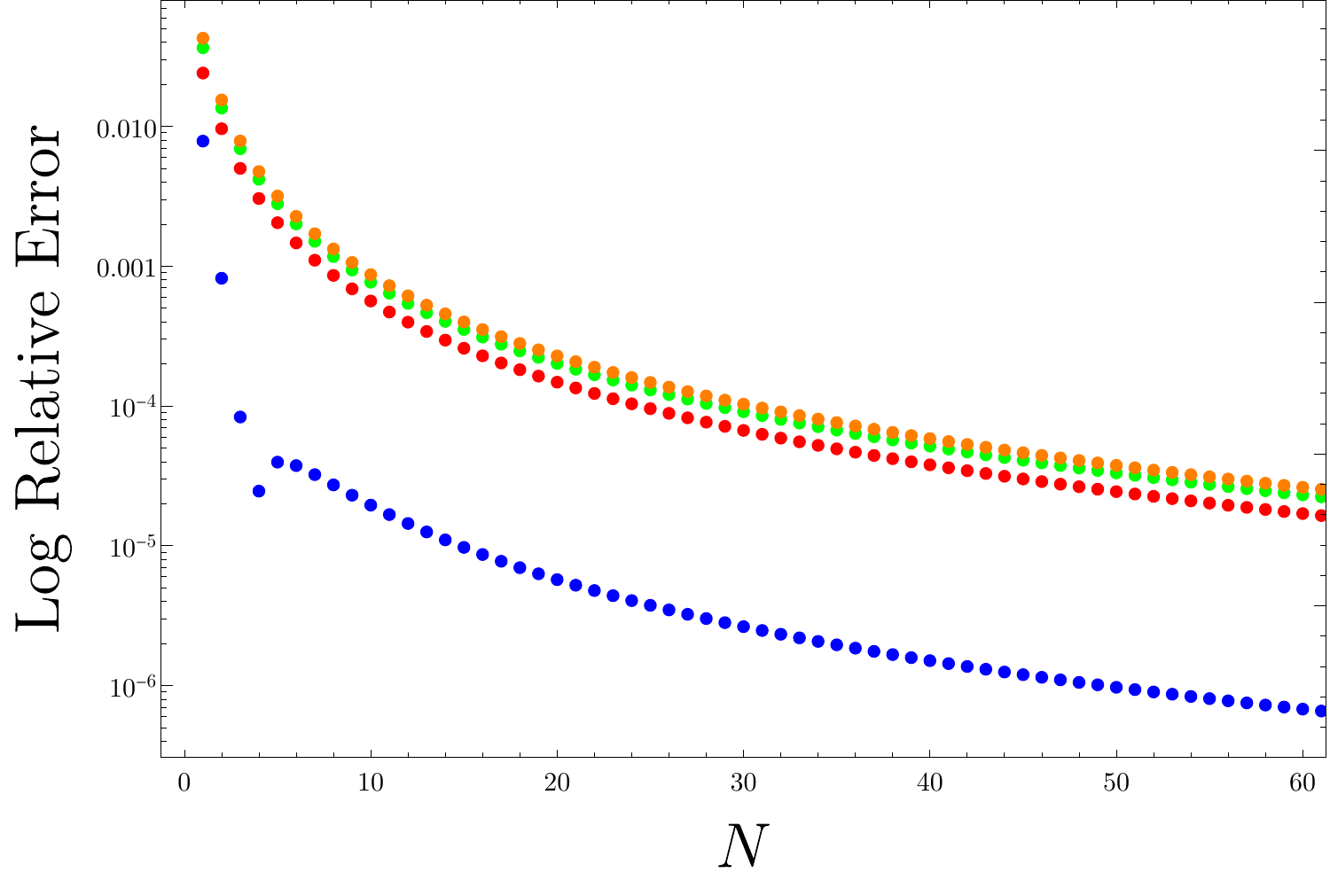}
\caption{A plot showing the error in the WKB approximation as a function of overtone number for $p = 3,4,5,6$ corresponding to the blue, red, green and orange dots, respectively. Here we show the logarithm of the relative absolute error, $|\omega_{\rm WKB} - \omega_{\rm Num}|/\omega_{\rm Num}$. }
\label{wkb_error}
\end{figure}

\begin{figure}
\centering
\includegraphics[width=0.9\linewidth]{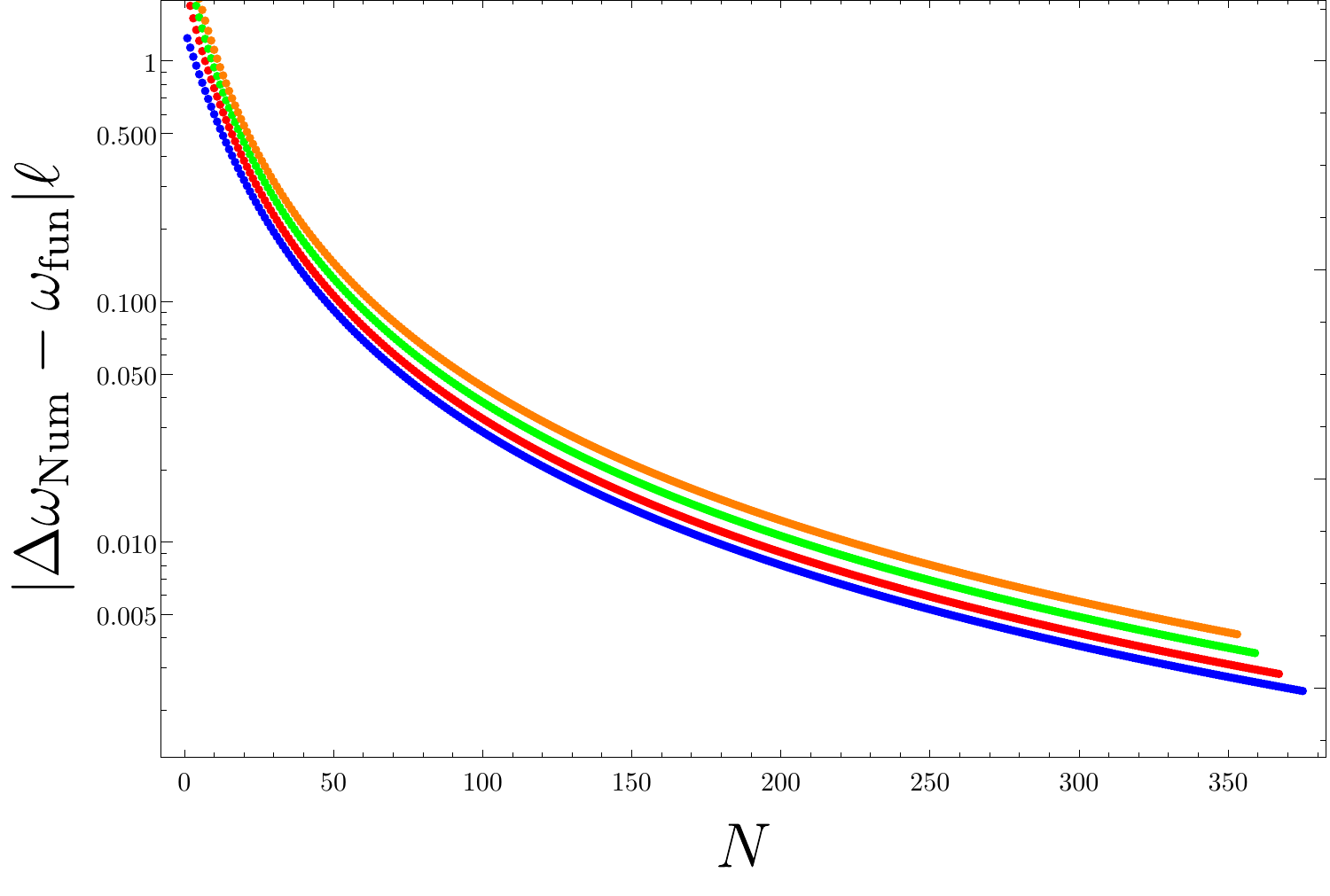}
\caption{A plot showing the difference between the overtone spacing for the numerically computed frequencies $\Delta \omega_{\rm Num} = \omega_{N+1} - \omega_{N}$ and the fundamental frequency $\omega_{\rm fun}$ on a logarithmic scale. The plot shows the case $n = 20, k =0$ for $p = 3,4,5,6$ corresponding to the blue, red, green and orange dots, respectively. The results are consistent with the fundamental frequency governing the spacing between successive overtones provided that $N \gg n, k$. }
\label{wkb_overtones}
\end{figure}

When confusion may arise, when referring to a particular element of $\{ \omega \}$ we will use the notation $\omega^{(p)}_{N, k, n}$ to indicate the dependence on the various parameters that appear in the equation. Here $N$ is a non-negative integer indicating the overtone, and $n = m/p$. Since the equations and boundary conditions are invariant under $m \to - m$, we will without loss of generality consider only the case where $n$ is a non-negative integer. When there is less risk of confusion, we will suppress additional data attached to the spectral element  $\omega^{(p)}_{N, k, n}$ to avoid unnecessary bulky notion. 

To begin our discussion of the numerical results, we compare the output of the pseudospectral method with the approximate results obtained via the WKB method. In the case of modes with $n = k = 0$, the WKB approximation is rather accurate for all overtone numbers. This is shown in Figure~\ref{wkb_compare}, where we show the output of the numerical methods (coloured dots) with the analytic approximation provided by the WKB method (black lines). The plot shows this comparison for overtones up to $N = 50$ for $p = 3,4,5$ and $6$. In all cases, the agreement between the WKB result and the more accurate numerical approach is superb. This is further backed up by Figure~\ref{wkb_error}, which shows the relative error between the WKB approximation and the numerical result for the same set of $p$ values. This plot shows that the WKB approximation is most accurate for the $p = 3$ case, but in all cases is accurate to about one part in one-thousand for overtone number $N \ge 10$. As expected, the accuracy of the WKB approximation becomes better the larger the overtone number becomes.

When $n$ and $k$ are non-vanishing, the WKB result no longer provides particularly good agreement, and we will discuss these cases at greater length below. However, it is important to note that even in these cases the WKB approach yields the correct spacing between overtones, provided the overtone number is sufficiently large compared to the values of $n$ and $k$. We illustrate this in Figure~\ref{wkb_overtones}, which compares the overtone spacing $\omega_{N+1} - \omega_{N}$ as obtained numerically to the fundamental frequency, which governs the overtone spacing in the WKB approximation. The plot shows this comparison as a function of overtone number for the $n = 20$ and $k=0$ mode for the cases of $p=3,4,5$ and $6$. We see in all cases that the absolute difference between the numerically determined spacing and the fundamental frequency is a monotonically decreasing function of the overtone number. This result, which we have verified in more examples in our numerical computations, is consistent with the notion that the fundamental frequency universally governs the spacing between overtones in the large overtone limit.

As mentioned, the WKB results provide a good approximation to the eigenvalues when $n = k = 0$, and more generally allow us to understand the spacing between overtones in the limit of large overtone number. However, there are a number of instances when the approximate solutions obtained in this way are insufficient. We move on to discuss these cases now.  While the Eguchi-Hanson-AdS$_5$ soliton is defined only for integer $p \ge 3$, the radial equation is sensible from a mathematical perspective under more general circumstances. To understand the spectrum $\{\omega\}$ for the soliton, we have found it useful to analyse the radial equation for real values of $p \in [2, \infty)$. The main reason for this is that it is possible, through a combination of numerical and analytical techniques, to obtain simple asymptotic forms for $\omega$ as $p \to 2$ and as $p \to \infty$.  By piecing together these approximate forms in various ways it is possible to get a good handle on the spectra over the full range of parameters.

\newpage
In the limit $p \to 2$ we find the following behaviour:
\begin{align}\label{smallpLim} 
\omega^{(p)}_{0, k, n} \approx& \, 2 \left(2 + k + n \right) + n \left(p-2 \right) 
\nonumber\\
&+ \left[\frac{6 \left(2n^2 - n - k \right)}{ (k+n)(1+2k+2n)(-1+2k+2n)} \right] (p-2)^2 + \cdots  
\end{align}
where the first term   has been obtained analytically; the analysis is outlined in Appendix~\ref{p2Sol}. Essentially, in the strict $p = 2$ limit, the radial equation reduces to that of a Klein-Gordon field on the orbifold AdS$_5/\mathbb{Z}_2$. That problem is obviously analytically solvable, and leads to the first term in the above. We find excellent numerical agreement between the $p\to 2$ limit of the numerical results and this analytically derived result, as detailed in Appendix~\ref{p2Sol}. 

The second and third terms in \eqref{smallpLim} have been inferred from the numerical results. For simplicity, we have focused here only on the fundamental mode, and have not worked out the dependence for the overtones due to the complexity. Our procedure to determine these corrections was the following. We have computed, for numerous choices of $n$ and $k$, the spectrum on the interval $p \in (2, 3]$. For a given choice of $n$ and $k$, we then fit the numerical results to a polynomial in $(p-2)$. Adjusting the order of the polynomial fit, we see that the coefficients converge rapidly to particular values\footnote{For example, including terms up to order $(p-2)^{10}$ in the fit we find that the coefficient of the linear term $(p-2)$ converges to five decimal places.}.  We take this as an indication that a series in $(p-2)$ captures accurately the behaviour of the spectrum in the close vicinity of $p=2$ for given values of $n$ and $k$. By comparing the results of this procedure across several different values of $n$ and $k$, we arrive at a collection of values $c_{n, k, i}$ for the coefficients of the $(p-2)^i$ term in the polynomial fit. We then study the way the different $c_{n,k,i}$ depend on $n$ and $k$, and from this infer their analytical dependence. The analytic dependence is then cross-checked against numerical results for values of $n$ and $k$ that were not part of the sample used when deducing the candidate analytic form. This process becomes more involved as the power $i$ is increased. However, for the linear and quadratic terms, the dependence is simple enough that it can be inferred, giving the results presented above\footnote{We have also found that the behaviour of the $(p-2)^3$ term can be deduced in certain limits. For example, we have found that for $n=0$ the coefficient of this term is $-3/(4k^2-1)$ while for $k= 0$ the coefficient of this term is $3/(2n+1)^2$. However, the functional form of the cubic term for general values of $n$ and $k$ has eluded us.}. The fact that the dependence of these terms  on $n$ and $k$ appears so simple suggests that it may be possible to determine these corrections analytically, though we shall not pursue this any further.
\begin{figure}
\centering
\includegraphics[width=0.45\textwidth]{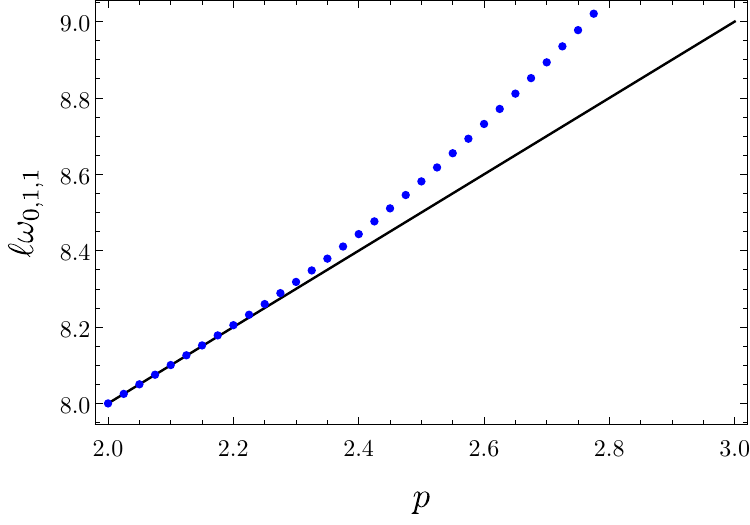} \quad
\includegraphics[width=0.45\textwidth]{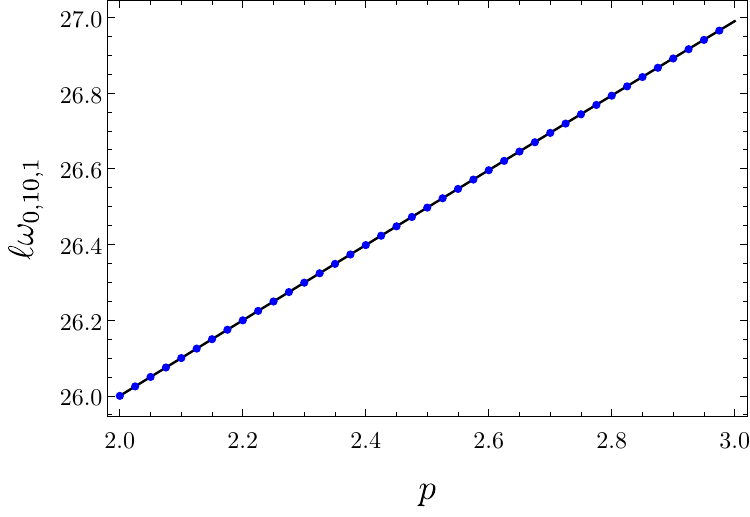}
\includegraphics[width=0.45\textwidth]{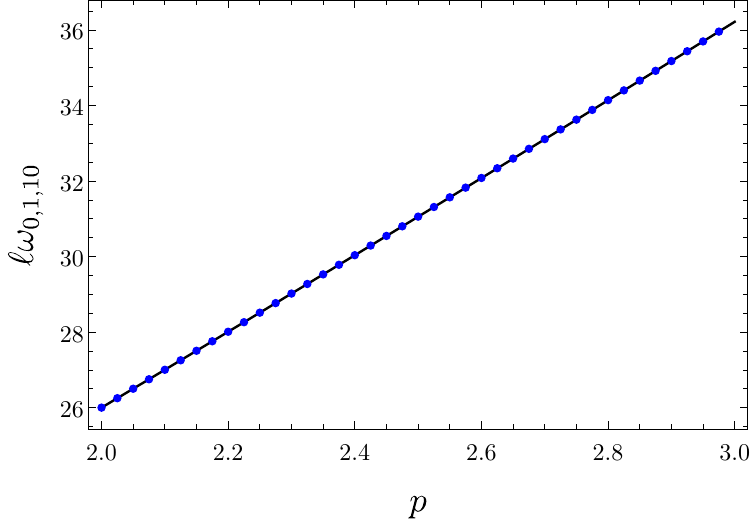} \quad 
\includegraphics[width=0.45\textwidth]{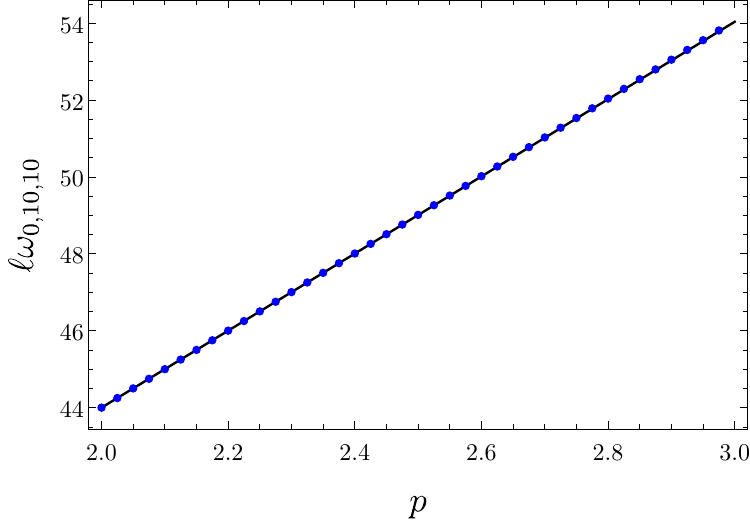}
\caption{Here we compare the approximate form of fundamental mode (black line) against the numerically determined spectra (blue dots) over the range $p \in (2, 3]$. The plots show the cases $n=1, k=1$ (top left), $n=1, k=10$ (top right), $n=10, k=1$ (bottom left) and $n=10, k=10$ (bottom right). The plots indicate that the approximate form is valid over a larger range of $p$ values as the value of $n$ or $k$ is made larger. }
\label{smallPCompare}
\end{figure}

In Figure~\ref{smallPCompare}, we compare the analytic approximation~\eqref{smallpLim} with the numerical results for different values of $n$ and $k$, over the interval $p \in [2, 3)$. While this range of $p$ does not correspond to regular five-dimensional geometries, it does concisely summarize important information about this approximation. First, we see that the approximate form is quite accurate over this interval. Second, we note that the approximation does better for large values of $n$ and $k$. The reason for this second fact seems to be the following. For large $n$ or $k$, the coefficient of the quadratic term in \eqref{smallpLim} behaves like $1/n$ or $1/k^2$, respectively. This suggests that series \eqref{smallpLim} --- if it is convergent --- converges more rapidly for large values of these quantum numbers, or --- if it is an asymptotic series rather than a convergent one --- that the series approximates the true function for a large range of $p$ values in this limit. 

\begin{figure}
\centering
\includegraphics[width=0.43\textwidth]{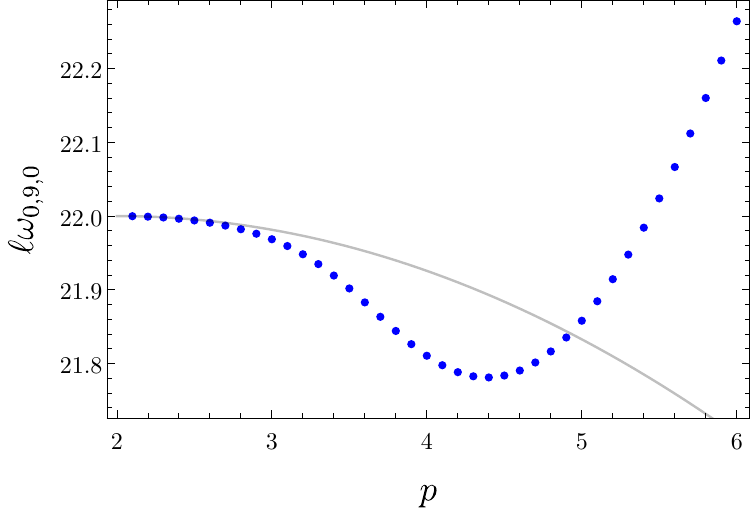}
\quad 
\includegraphics[width=0.43\textwidth]{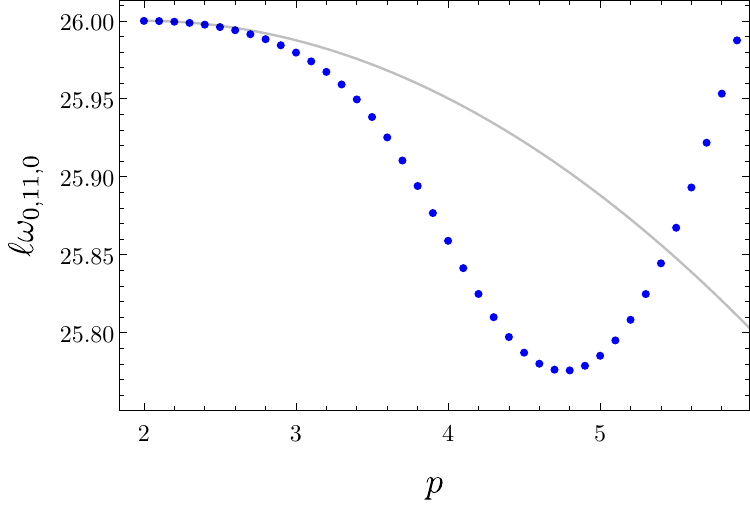}
\includegraphics[width=0.43\textwidth]{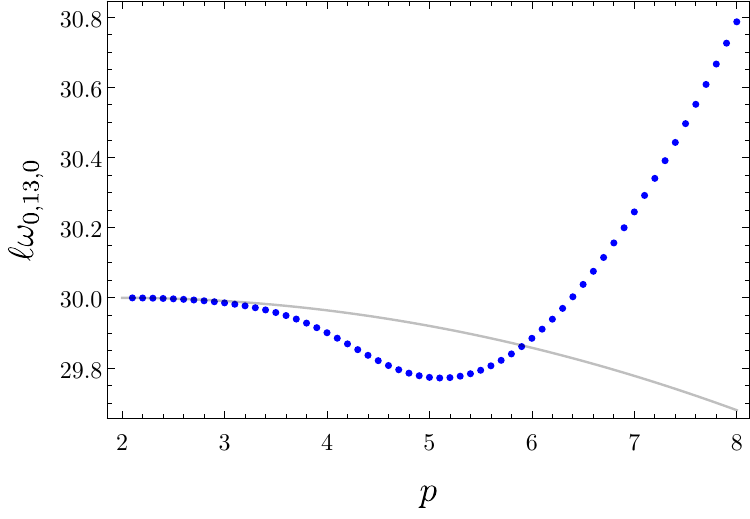}
\quad 
\includegraphics[width=0.43\textwidth]{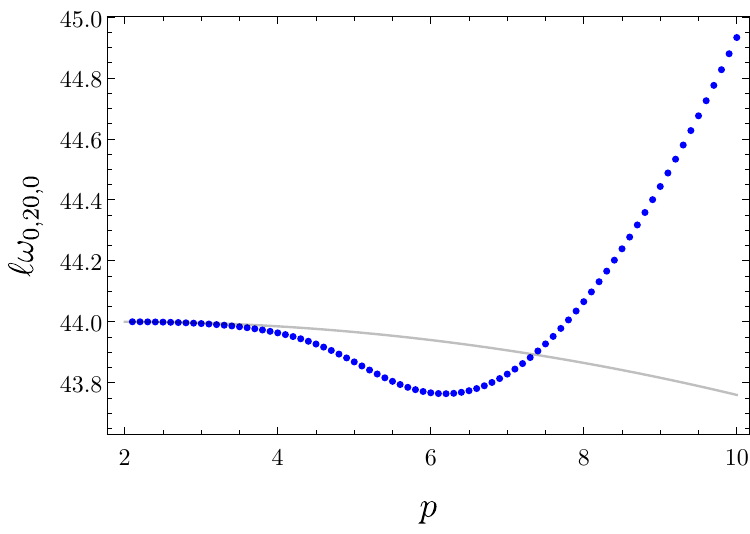}
\caption{Plots of the frequencies vs.~$p$ for $n=0$ and $k = 9, 11, 13$ and $20$ (top left, top right, bottom left, bottom right, respectively). These plots highlight the fact that, for $k > 2n^2 - n$, it is possible for the eigenvalue to initially decrease as $p$ increases. In all cases, the light gray line indicates the analytic approximation~\eqref{smallpLim}, while the blue dots indicate numerical data.}
\label{curvy}
\end{figure}

A careful analysis of the analytic approximation shows that the quadratic term is \textit{negative} whenever $ k> 2n^2-n$. This leads to some interesting structure in the spectrum, but does not lead to any complex eigenvalues. Namely, we find that in some circumstances the eigenvalues can initially \textit{decrease} as a function of $p$, however this behaviour is ultimately reversed when $p$ becomes large --- for sufficiently large $p$, the eigenvalues are monotonically increasing as a function of $p$, as we will discuss below.  Examples are shown in Figure~\ref{curvy} for $n=0$ and different choices of $k$. For all values of $k$ that we have explored, the shape of the eigenvalue curve as a function of $p$ is qualitatively the same. At larger values of $k$, the `dip' occurs at larger values of $p$. In all cases, the depth of the dip is rather small, and in no cases do the eigenvalues come close to crossing through zero. In practice, we observe this effect for physical values of $p$ only in the case $n=0$. For $n \neq 0$, the small $p \to 2$ approximation is less accurate for $k > 2n^2-n$. Indeed, for any fixed values of $n$ and $k$, the approximate form derived in the $p\to 2$ limit will ultimately become bad when $p$ is sufficiently large. Roughly, this occurs when the quadratic piece in the approximation dominates.

When $p$ becomes sufficiently large compared to the quantum numbers $n$ and $k$, the approximate form derived in the $p \to 2$ limit fails to accurately capture the details of the spectrum. In this regime, we can make progress by understanding the solution of the radial equation in the limit of large $p$. Numerical experiments suggest that, when $p \gg n, k$ the eigenvalues exhibit a linear dependence on $p$. This observation can be explained analytically. As described in Section~\ref{largeP}, in the large $p$ limit, there is a formal sense in which the geometry limits to the AdS soliton. The normal modes of the AdS soliton then govern the slope of the eigenvalues $\omega_{N, k, n}$. Explicitly, we have the following relation that holds for large $p$:
\be\label{bigPomg} 
\frac{d \omega^{(p)}_{N , k , n}}{d p} = \alpha_{N, n} + \dots  \quad \text{as} \quad p \to \infty \, ,
\ee
where the dots denote terms that are subleading as $p \to \infty$. Here $\alpha_{N, n}$ are the normal modes of the AdS soliton which, unfortunately, cannot be determined analytically\footnote{Strictly speaking, $\alpha_{N, n}$ are the normal modes for the AdS soliton when the time coordinate is rescaled by a factor of $2$ from the usual value.}.  We tabulate in Appendix~\ref{largePApp} the numerically determined values of these normal modes and overtones for several values of $n$. We also note that the WKB approximation gives a reasonably accurate approximation when $N \gg n$,
\be 
\ell \alpha_{N, n} \approx \frac{2 \sqrt{2} \pi^{3/2}}{\Gamma[1/4]^2} \left(N + \frac{1}{2} \right) \, .
\ee
Note that the fact that the quantum number $k$ is absent in the above expressions is result of neither an assumption nor a typo. There is an `emergent degeneracy' that appears for sufficiently large $p$ that eliminates any dependence on $k$ in the leading expressions.

\begin{figure}
\centering
\includegraphics[width=0.45\linewidth]{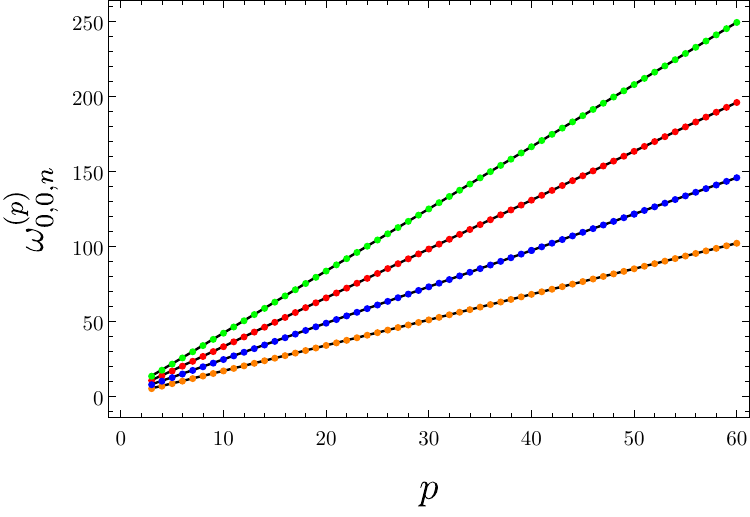}
\quad 
\includegraphics[width=0.45\linewidth]{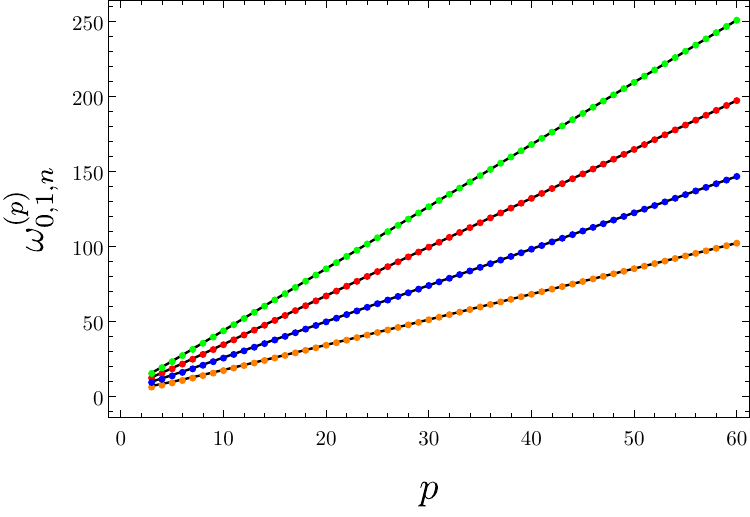}
\quad
\includegraphics[width=0.45\linewidth]{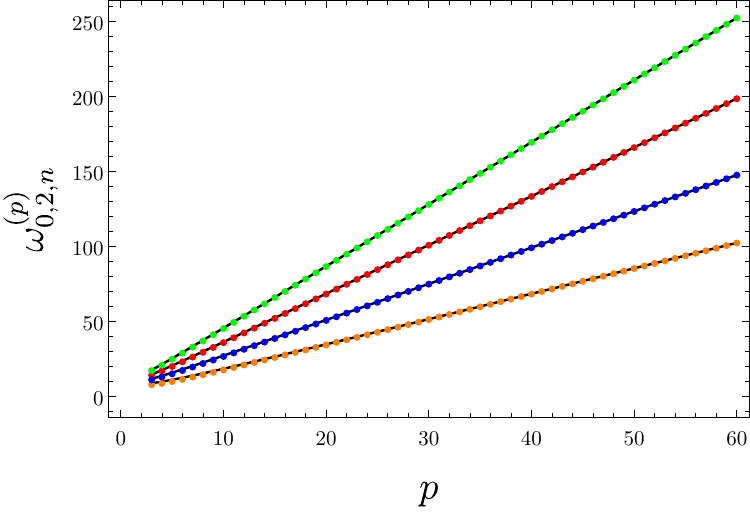}
\caption{Plot of the fundamental mode for $k=0$ (top left), $k=1$ (top right), and $k=2$ (bottom). In all cases, the curves correspond to $n=0,1,2,3$ in order of bottom to top (or, in colour order, orange, blue, red, and green). The solid black lines correspond to an analytic approximation to the curves, while the coloured discs correspond to the results of the numerical computation.}
\label{bigPfig}
\end{figure}

While~\eqref{bigPomg} provides a good approximation for the slope, additional details are required to determine the intercept and thereby obtain a good approximation to the eigenvalues themselves, rather than just their slope. One simple --- and surprisingly accurate --- option is to combine this large $p$ approximation with the fact that we know as $p \to 2$ the eigenvalues approach $2(2 + n +k + N)$. Choosing the constant of integration to be such that the eigenvalues have this point as the intercept, we obtain the approximation:
\be\label{bigPapprox} 
\omega^{(p)}_{N , k , n} \approx (p-2) \alpha_{N, n} + 2(2+n+k+N) \, .
\ee
This approximate form gives very good agreement with the numerically determined eigenvalues over the full range of physical $p$ values, provided that $n$ is sufficiently large. In this sense, one may consider (to a first approximation) that the normal modes of Eguchi-Hanson-AdS interpolate between the normal modes of AdS$_5/\mathbb{Z}_2$ and the AdS soliton as the parameter $p$ is varied. 

A more accurate approximation can be obtained by carrying out the same series of steps that we did in the $p\to2$ limit. That is, through a combination of numeric and analytic techniques, we can determine the next-to-leading order terms in the large $p$ expansion. If instead we do this, we obtain the following:
\be 
\omega^{(p)}_{N , k , n} \approx p \alpha_{N, n} + \frac{n(2k+1)}{\alpha_{N, n}} + \mathcal{O}(1/p) \, .
\ee
We show this approximation for a few representative cases in Figure~\ref{bigPfig} for small values of $n$.   

Despite the constant term being heuristically derived, the approximation \eqref{bigPapprox} actually does a better job than one might naively expect, at least for the fundamental mode. One reason for this is the following. Though we have not been able to prove this behaviour analytically, we numerically observe that as $n$ becomes large, $\alpha_{0, n} \approx n$. Therefore, in this limit, the large $p$ approximation given in~\eqref{bigPapprox} approaches
\be 
\omega^{(p)}_{0 , k , n} \approx 2(2+n+k) + n (p-2) \quad \text{for} \quad n \to \infty \, .
\ee
If we compare this result with the large $n \to \infty$ limit of the $p\to 2$ approximation given in~\eqref{smallpLim} we see that the two expressions are exactly the same. Therefore, the approximate forms derived in the $p \ll n$ and $p \gg n$ limits are identical. This explains why both approximate forms do better than expected for large values of $n$.

For the sake of completeness, we tabulate a number of the numerically determined normal modes in Appendix~\ref{modeTables}.


\section{Conclusions}
\label{sec6}

Eguchi-Hanson-AdS solitons are conjectured to be the ground states for anti de Sitter gravity with lens space $L(p,1)$ boundary  conditions at infinity, analogous to how the AdS soliton is the conjectured ground state for toroidal boundary conditions at infinity. In this manuscript, we have considered various aspects of Eguchi-Hanson-AdS$_5$ solitons. Our primary objective, which we have initiated here, is to understand to what extent these geometries are likely to be stable or not. In this sense, it is known~\cite{Dold:2017hwr} that under certain circumstances perturbations of the geometry may result in the formation of naked singularities. However, the precise mechanism underlying such an instability remains unknown. Since relatively little is understood about these geometries, our study here has focused on other related properties as well.

A key observation we have made concerns the relationship between the Eguchi-Hanson solitons and the AdS soliton. We demonstrated that in the limit where the lens space $L(p,1)$ parameter $p \to \infty$, the geometry becomes identically equal to that of the AdS soliton, up to a rescaling of the time coordinate. This observation is more than a curiosity, as we have found here this allows one to obtain approximations for various quantities of interest in terms of those same quantities for the simpler AdS soliton geometry.\footnote{After the first version of this manuscript appeared, Edgar Shaghoulian brought to our attention~\cite{Shaghoulian:2016gol}, where the same observation was made for the partition functions of CFTs on lens spaces. Our result for the connection between Eguchi-Hanson-AdS and the AdS soliton could be inferred from these results.} 

We have studied the conserved quantities of the Eguchi-Hanson solitons and constructed the extended first law of soliton mechanics. In doing so, we observe that these solitons possess a non-trivial thermodynamic volume, despite having vanishing entropy. Commensurate with a previous study of thermdynamic volume for asymptotically globally AdS$_5$  solitons~\cite{Andrews:2019hvq}, the source of this quantity is topological in nature, and has to do with the structure of the Killing potential in the vicinity of the bubble. It would be interesting to extend these observations for other examples of solitons, as it may allow for further elucidation of the role and interpretation of thermodynamic volume in gravitational thermodynamics. 

We also considered the thermodynamics of Eguchi-Hanson solitons in the canonical ensemble. After computing the on-shell action, we showed that for low temperatures the soliton dominates the canonical ensemble, while at higher temperature a Schwarzschild-AdS-type black hole with lens space horizon dominates. The phase transition between these two states is of the same type as studied for the toroidal AdS black hole and the AdS soliton first studied in~\cite{Surya:2001vj}.

We examined the geodesics of the Eguchi-Hanson solitons, finding no evidence suggestive of instability such as trapping of null geodesics. However, the time-like geodesics exhibit are oscillatory in nature analogous to geodesics in global AdS. Thus, an instability of the type shown in \cite{Bizon:2011gg} for pure AdS may be present also for these solitons. Investigating this in more detail is something we hope to return to in the future. 

Finally, we studied the separable solutions to the massless Klein-Gordon equation on this background. Via numerical and approximate analytical methods, we find a set of modes that oscillate, never decaying, analogous to AdS. We find no evidence of frequencies with a non-vanishing imaginary component. Using the WKB approximation, we have been able to study the separation between successive overtones, showing that it is well-approximated by the light-crossing time of the geometry by radial null geodesics. We have found it possible to understand the normal mode frequencies by piecing together two approximations. The first involves an analysis of the equation in the vicinity of the value $p \to 2$. While this does not correspond to a physical soliton solution, the radial equation in this limit reduces to that for AdS$_5 / \mathbb{Z}_2$, which admits analytic solutions. The second approximation involves the $p \to \infty$ limit of the radial equation, which reduces to the wave equation on the AdS soliton. Joining the two approximations together gives a quite accurate approximation for the normal modes of the Eguchi-Hanson solitons for any physical value of $p$ in terms of a single parameter that must be numerically determined: the normal modes for the AdS soliton. In an approximate sense, one can then consider the normal modes of the Eguchi-Hanson solitons as interpolating between those of 
the orbifold AdS$_5 / \mathbb{Z}_2$ and the AdS soliton.

There remain a number of areas for future investigation. First among these, along the lines of understanding potential mechanisms for instability, would be an understanding of the gravitational perturbations of the geometry. This task is simplified since the solution is cohomogeneity-one, allowing for the techniques of~\cite{Murata:2008yx} to be used. An understanding of these perturbations may hint toward the existence of `resonating' solutions constructible as non-linear extensions of the normal mode solutions~\cite{Ishii:2020muv, Garbiso:2020dys}. It would furthermore be interesting to understand the implications of these geometries within the AdS/CFT correspondence, where they may be relevant for understanding aspects of confined phases of CFTs on lens space geometries~\cite{Constable:1999gb, Myers:1999psa, Myers:2017sxr}. It would also be interesting to understand the role of higher-curvature corrections for these geometries~\cite{Wong:2011aa, Corral:2021xsu, Corral:2022udb}.

\section*{Acknowledgements}
We would like to thank Roberto Emparan, Edgar Shaghoulian, and Eric Woolgar for helpful discussions, comments, and correspondence. The work of RAH received the support of a fellowship from ``la Caixa” Foundation (ID 100010434) and from the European Union’s Horizon 2020 research and innovation programme under the Marie Skłodowska-Curie grant agreement No 847648” under fellowship code LCF/BQ/PI21/11830027. HKK acknowledges the support of the NSERC Grant RGPIN-2018-04887. The work of TD and RBM was supported by the Natural Sciences and Engineering Research Council of Canada.

The lands on which Memorial University’s campuses are situated are in the traditional territories of the Beothuk, Mi’kmaq, Innu, and Inuit of the province of Newfoundland and Labrador. McMaster University is located on the traditional territories of the Mississauga and Haudenosaunee nations and within the lands protected by the “Dish with One Spoon” wampum agreement. University of Waterloo is situated on the traditional territory of the Neutral, Anishinaabeg and Haudenosaunee peoples. 
\appendix

\section{Solution of the Radial Equation via the Shooting Method}
\label{shooting}

In this section, we provide independent confirmation for the results obtained via the pseudospectral method. To this end, we consider the solution of the radial equation via the shooting method. 

We construct two numerical solutions of the radial equation. One solution begins near the bubble at $r = a$, using a power series solution of the differential equation in this neighbourhood to construct initial conditions, along with an initial guess for the eigenvalue $\omega$. This solution is integrated outward, toward $r = \infty$, using the standard built-in numerical methods of Mathematica. The second numerical solution proceeds in much the same way, but begins near the asymptotic boundary, and is integrated inward toward $r = a$. 

The idea is simple: if $\omega$ has been chosen correctly, then the solution must be everywhere regular, and moreover, the two numerical solutions should agree over the domain. However, since the equation is linear, and both $R(r)$ and $c R(r)$ for some constant $c$ are equally valid solutions, it is not so simple to directly compare the solutions obtained by integrating from either end. (As there is no simple way to ensure consistent normalization from the different starting points.) Instead, we compare a logarithmic derivative,
\be 
\frac{R'_{\rm in/out}}{R_{\rm in/out}}\,,
\ee
of the numerical solutions at intermediate points. In this way, differences in the normalization of the solution can be eliminated.  Matching of the logarithmic derivatives of the two numerically constructed solutions over the domain is taken to mean the correct value of $\omega$ has been identified.

As the purpose of this method is to serve as independent confirmation of the pseudospectral results, we will not perform an exhaustive analysis here. Rather, we will present a few instances that demonstrate consistency of the two approaches. 

\begin{figure}
\centering
\includegraphics[width=0.3\textwidth]{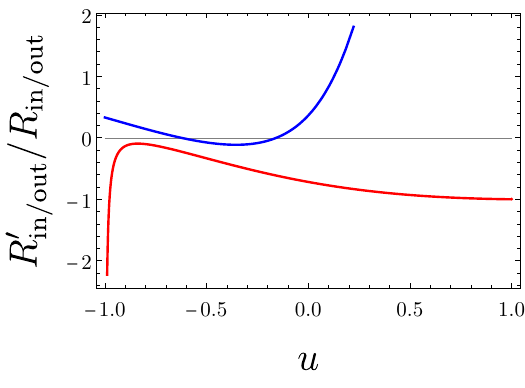}
\quad
\includegraphics[width=0.3\textwidth]{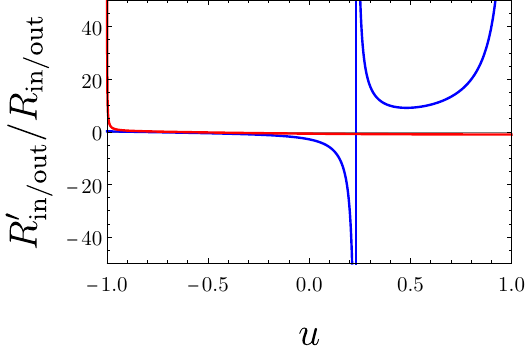}
\quad
\includegraphics[width=0.3\textwidth]{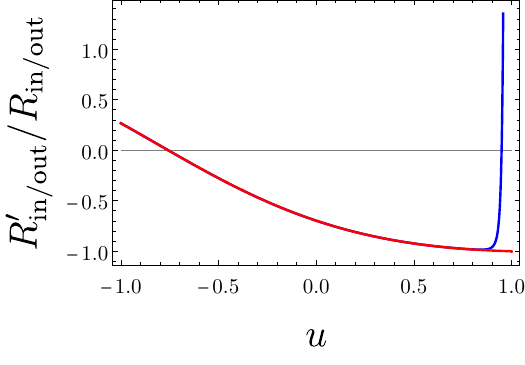}
\caption{Plots of the logarithmic derivative of the numerical solutions. The blue curves correspond to solutions integrated from $u=-1$, while the red curves correspond to solutions integrated from $u=+1$. In this case, we have set $p = 3$, $n=0$, $k=0$.  The plots correspond to $\omega = 5.2, 5.4, 5.2985999$ in order of left to right. }
\label{shooting_logs}
\end{figure}

We show in Figure~\ref{shooting_logs} an example of how the process works. The graphs show the logarithmic derivative of the solution over the domain $u \in [-1, 1]$. The blue curves are those solutions obtained by numerically integrating the equation beginning at $u = -1$, while the red curves are those solutions obtained by numerically integrating the equation beginning from $u = +1$. The three graphs correspond to three different choices of the eigenvalue $\omega$, while we have set $p = 3$, $n=0$ and $k=0$ here. The pseudospectral method outputs a value $\omega_{\rm PS} = 5.2985999$ --- this choice of $\omega$ appears in the rightmost graph. The leftmost graph corresponds to the choice $\omega = 5.2$, while the center one corresponds to $\omega = 5.4$. For both cases shown where $\omega$ is different from $\omega_{\rm PS}$, the two solutions are clearly in disagreement. However, for the choice $\omega = \omega_{\rm PS}$ the curves are visually indistinguishable over much of the domain, except for very near the end points\footnote{Quantitatively, the difference in solutions is on the order of $10^{-6}$ over the interval $u \in [-1/2, 1/2]$. }. This is consistent with $\omega = \omega_{\rm PS}$ being the correct choice for a consistent solution. 

It is also insightful to understand the difference in the in/out solutions as a function of $\omega$, to ensure that the implementation of the pseudospectral method is not missing certain overtones, or returning incorrect eigenvalues. To this end, it is useful to introduce an integrated residual. Of course, for any finite resolution, we expect that the numerical solution that begins from $u = -1$ to blow up sufficiently close to $u = +1$, and vice versa for the solution that begins from $u = +1$. Therefore, an integrated residual should focus on an intermediate domain where both solutions can be expected to be accurate. For this, we (arbitrarily) choose $u \in [-1/2, 1/2]$. Another issue is the following. When $\omega$ becomes larger than the fundamental frequency, the solution possesses a zero somewhere in the domain. (The number of zeros increases as $\omega$ becomes larger than the various overtones). This leads to poles in the logarithmic derivative --- see the center plot of Figure~\ref{shooting_logs} for an example.  To remedy this, it is convenient to work with an integrated \textit{reciprocal} residual,
\be 
{\rm Res}(\omega) = \int_{-\frac{1}{2}}^{\frac{1}{2}} \frac{du}{\left| \partial_u \log R_{\rm out} - \partial_u  \log R_{\rm in} \right|} \, .
\ee 
Viewed as a function of $\omega$, the residual ${\rm Res}(\omega)$ will peak when the difference between the two solutions tends to zero. We can identify those peaks as the values of $\omega$ where a consistent solution exists. 

\begin{figure}
\centering
\includegraphics[width=0.45\textwidth]{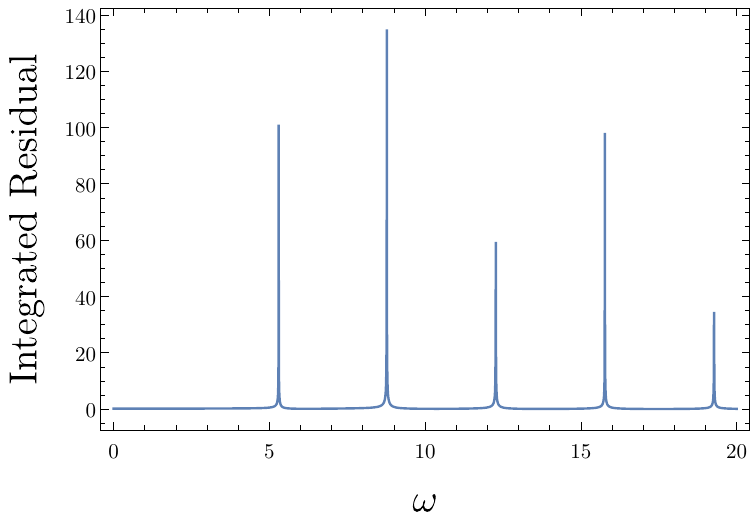}\quad
\includegraphics[width=0.45\textwidth]{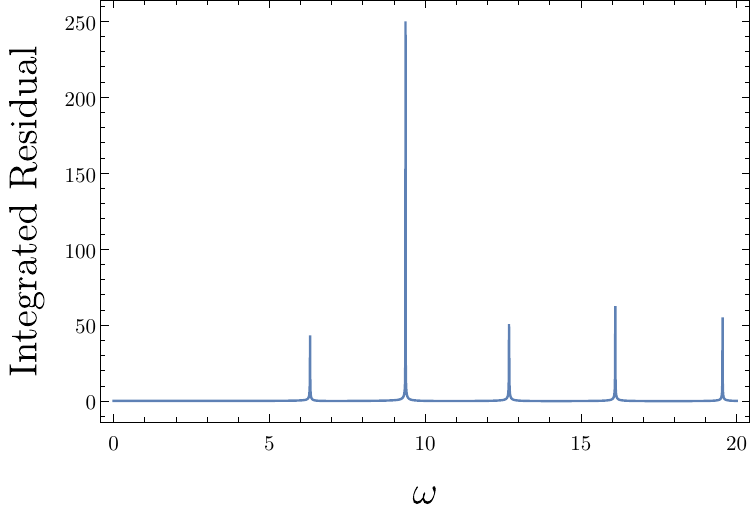}
\includegraphics[width=0.45\textwidth]{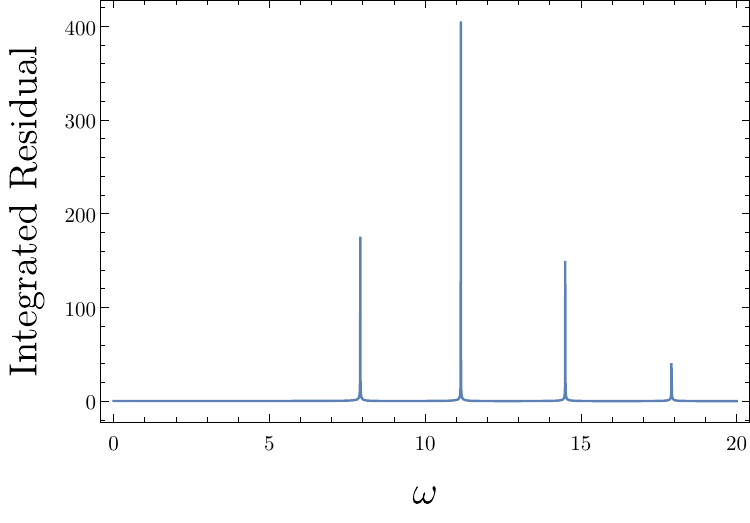}\quad 
\includegraphics[width=0.45\textwidth]{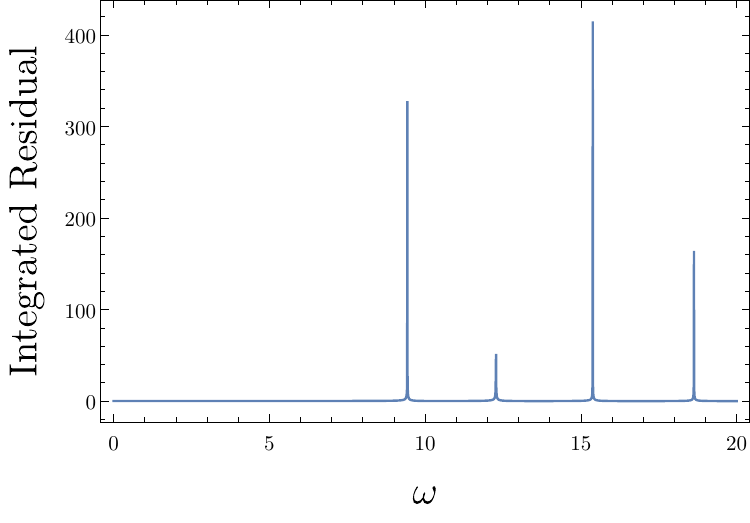}
\includegraphics[width=0.45\textwidth]{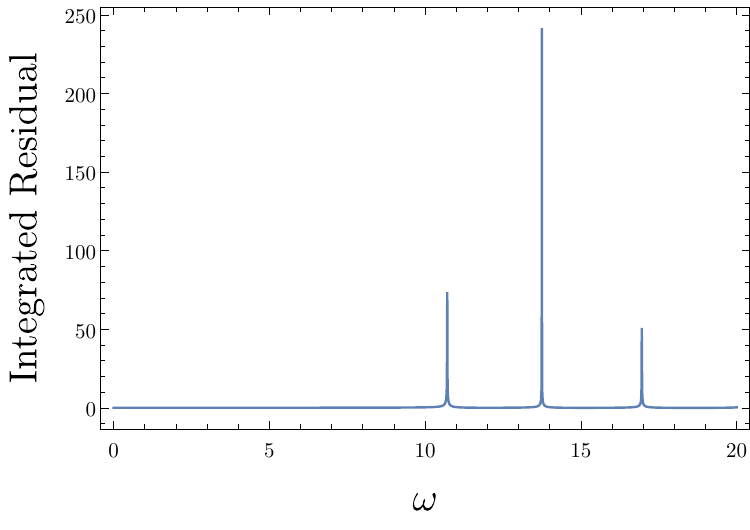}\quad
\includegraphics[width=0.45\textwidth]{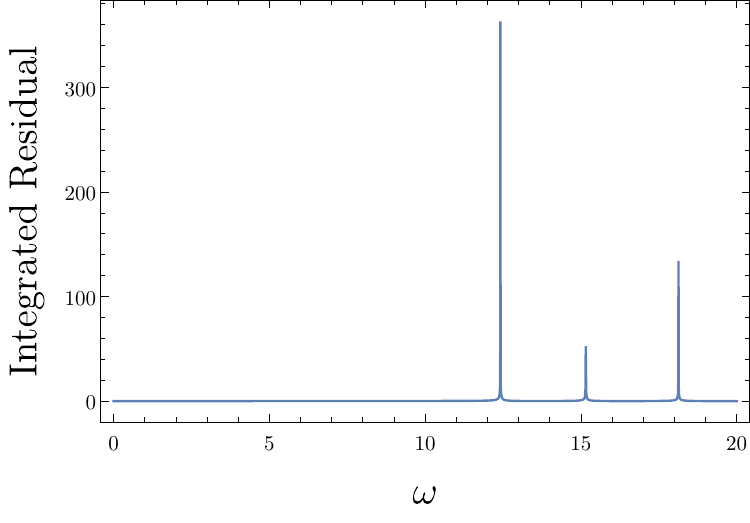}
\caption{Plots of the integrated (reciprocal) residual for $n = 1,2,3$ (first, second, and third rows, respectively) and $k=0, 1$ (first and second columns, respectively). In all cases, $p=3$ has been chosen. The reciprocal residual peaks at the values of $\omega$ for which a consistent solution exists --- the absolute heights of the peaks shown in the graphs is meaningless. In all cases, the peaks correspond precisely to those values of $\omega$ determined via the pseudospectral method.} 
\label{residual}
\end{figure}

We show a few representative examples in Figure~\ref{residual}. In these plots, the peaks correspond to values of $\omega$ for which the boundary value problem admits a sensible solution. In all cases, these peaks correspond exactly to those values of $\omega$ obtained via the spectral method. No additional values of $\omega$ have been obtained, and no errors have be determined.

\section{Radial Equation at Large $p$: Extracting the Slope of the Eigenvalues}
\label{largePApp}
As discussed in the numerical solution of the radial equation, it is fruitful to consider the radial equation for arbitrary values of $p$. In particular, we have discussed the numerical observation that the behaviour of the eigenvalue $\omega$ rapidly becomes dominated at large $p$ by a linear dependence. This fact can be explored in more detail, semi-analytically, to give a better understanding of that result. 

Consider the radial equation~\eqref{radialEq} and perform the transformation
\be 
r = z a \, ,
\ee
which maps the domain to $z \in [1, \infty)$. After this transformation, the resulting equation is given by
\begin{align}\label{largep1}
0 =& z (z^4-1)\left(4 + (p^2-4)z^2\right) R''_p(z) + \left(4 -(p^2-4) z^2 + 12 z^4 + 5  (p^2-4) z^6 \right) R'_p(z) 
\nonumber\\
&+ 4 z^3 \left[ -4 k^2 - 2 n p - 4 k (1+np) + \frac{n^2 p^2 z^4}{1-z^4} + \frac{(p^2-4) z^2 \ell^2 \omega^2}{4 + (p^2 -4)z^2}\right] R_p(z) \, ,
\end{align}
where we have included the subscript $p$ to reinforce that, here, we are thinking of this as an arbitrary constant parametrizing the solution. 

Consider~\eqref{largep1} in the limit where $p$ is very large. Doing so, it becomes clear that the leading-order behaviour in each term behaves as $\mathcal{O}(p^2)$. Thus, a consistent possibility is for the dominant $p$-dependence of $\omega$ to be such that 
\be 
\omega \sim \frac{\alpha}{\ell} p
\ee
at large $p$. 

We can extract the equation that governs this behaviour and determines $\alpha$. Rescale $\omega = \alpha p$ and peel off the terms in the differential equation that behave as $\mathcal{O}(p^2)$ at large $p$. For consistency of notation, let us call $R_p(z)$ in this limit $h_0(z)$. Carrying this out, we find the following differential equation:
\be 
0 = (-1+z^4) h''_0(z)  + \frac{(-1+5 z^4)}{z} h'_0(z) + 4 \left(\frac{n^2 z^4}{1-z^4} + \alpha^2 \right) h_0(z) \, .
\ee

By determining the values of $\alpha$ for which regular solutions to the above equation exist, we can determine the large $p$ slope of the eigenvalues $\omega$. Note that it is only the quantum number $n$ $(=m/p)$ that enters into the equation, as $k$ has no fixed dependence on $p$.  Thus, the eigenvalues $\alpha = \alpha_{N, n}$ are characterized by two numbers: the overtone $N$ and the quantum number $n$. In other words, there is an `emergent degeneracy' at large $p$: The eigenvalues for all choices of $k$ are effectively the same, provided that $p > k$ is sufficiently large.

Unfortunately, we have not been able to solve this problem analytically (the equation is of Heun type). However, it is straight-forward to apply the same pseudospectral numerical techniques to this problem. We do not repeat the basic set up of this problem, and proceed directly to the results. 

\begin{table}
\begin{center}
\begin{tabular}{c c c c} 
 \toprule
 \multicolumn{4}{c}{Values of $\alpha_{N, n}$} \\
 \toprule
 $n$ &  \multicolumn{3}{c}{$N$} \\ \cline{2-4} 
  & 0 & 1 & 2 \\
 \hline\hline
 0 & 1.70203 & 2.93798 & 4.15256 \\
 1 & 2.42243 & 3.61436 & 4.80816 \\
 2 & 3.25439 & 4.38480 & 5.54238 \\
 3 & 4.14123 & 5.21301 & 6.33085 \\
 4 & 5.05869 & 6.07883 & 7.15802 \\
 5 & 5.99504 & 6.97048 & 8.01368 \\
 6 & 6.94397 & 7.88064 & 8.89089 \\
 7 & 7.90176 & 8.80456 & 9.78483 \\
 8 & 8.86607 & 9.73904 & 10.69202 \\
 9 & 9.83533 & 10.68180 & 11.60991 \\
 10 & 10.80848 & 11.63122 & 12.53658\\
 \hline \hline
 50 & 50.47989 & 50.98209 & 51.54973 \\
 100 & 100.38136 & 100.78145 & 101.23488 \\
 500 & 500.22318 & 500.45767 & 500.72385 \\
 1000 & 1000.17715 & 1000.36329 & 1000.57462 \\ 
 \bottomrule
\end{tabular}
\end{center}
\caption{A selection of numerically obtained  $\alpha_{N, n}$ values.}
\label{alphaTab}
\end{table}

Some selected values of $\alpha_{N,n}$ are presented in Table~\ref{alphaTab}. These have been computed using the pseudospectral method to the accuracy shown in the table. While there is not a clear, discernible pattern for smaller values of $n$, it is clear that when $n$ becomes large we have $\alpha_{N, n} \approx n$. Unfortunately, it becomes increasingly expensive to perform the computations when $n$ becomes very large. Computing $\alpha_{N, n}$ to five decimal places requires a Chebychev discretization of more than 1000 points when $n$ exceeds about 1000. 

\subsection{Subleading Terms in a Large $p$ Expansion}

In the previous subsection, we have extracted a differential equation whose solution leads to the leading-order slopes of the eigenvalues $\omega$ at large values of $p$. It is possible to do somewhat better than this, without encountering additional obstructions. To this end, we approach the problem as one in perturbation theory, expanding in the small parameter $\rho = 1/p$. 

We begin by expanding the eigenvalue and the eigenfunction as a perturbative series in $
\rho$:
\begin{align}
\rho^2 R_p(z) &= h_0(z) + \rho h_1(z) + \rho^2 h_2(z) + \cdots \, ,
\\
\rho^2 \omega^2 &= \alpha_{N, n}^2 + \rho \beta_1 + \rho^2 \beta_2 + \cdots \, .
\end{align}
These expansions are then inserted into the differential equation, and an expansion in powers of $\rho$ is performed. In general, determining the corrections requires knowledge of both the eigenvalues and the eigenfunctions at previous orders. However, a happy accident at $\mathcal{O}(\rho)$ allows for direct determination of $\beta_1$. The differential operator for $h_1(z)$ is identical to that determining $h_0(z)$, allowing for all terms involving derivatives to be eliminated. The problem then reduces to an algebraic equation that determines $\beta_1$, with the result
\be 
\beta_1 = 2n(1+2k) \, .
\ee

Unfortunately, at higher orders it does not appear to be possible to obtain analytic results for the corrections. While progress could be made numerically, the problem becomes more complicated. Therefore, it is not clear there is any benefit to pursuing this path further. 

\section{Eguchi-Hanson Solitons as $p \to 2$}
\label{p2Sol}

Having fruitfully studied the large $p$ limit of the Eguchi-Hanson soliton in the previous appendix, here we will consider the limit $p\to 2$. The limit should be taken at fixed cosmological scale $\ell$, so that the theory under consideration is not altered. To this end, we write $a = \mathscr{A} \ell$ and consider the limit $p \to 2$. This is simple, and just results in $\mathscr{A} \to 0$. The resulting geometry is
\begin{align}
ds^2 &= - g(r) dt^2 + \frac{dr^2}{ g(r)}
+ \frac{r^2}{4}  \left[d\psi + \cos(\theta) d\phi\right]^2
+ \frac{r^2}{4} d\Omega^2_2 \, , \quad
g(r) = 1 + \frac{r^2}{\ell^2} \, ,
\end{align}
with $\psi$ normalized to have period $2 \pi$. The metric on the sections of constant $(t,r)$ is that of the projective space $\mathbb{S}^3/\mathbb{Z}_2$. The analysis of the wave equation on this space is useful for understanding the normal modes of Eguchi-Hanson-AdS for smaller values of $p$.

\subsection{Solution of the Radial Equation as $p \to 2$}

While (regular) Eguchi-Hanson solitons exist only for integer $p \ge 3$, it is sensible from the mathematical point of view to study properties of the radial equation~\eqref{radialEq} without this restriction imposed on $p$. Within this line of thought, the case $p=2$ is special, since for $p=2$ the parameter $a$ vanishes. That is, if we directly substitute $p=2$ into the radial equation, it reduces to the radial equation for a scalar field on AdS, but with a special choice $m = 2n$ inherited from the fact that the limiting geometry is not globally AdS$_5$, but instead the quotient space AdS$_5$/$\mathbb{Z}_2$. In this special case, the equation can be solved directly. 

Substituting $p=2$ into the radial equation we obtain,
\be 
0 = r^2(r^2+ L^2) R''(r) + r (3 L^2 + 5 r^2)R'(r) + L^2 \left(-4(k+n)(1+k+n) + \frac{L^2 r^2 \omega^2}{L^2 +r^2} \right)R(r) \, .
\ee
After setting $L=1$, the above equation has the following solution in terms of hypergeometric functions
\begin{align} 
R(r) =& \, (1+r^2)^{\omega/2} \left[C_1 r^{2(-1-k-n)} {}_2F_1 \left(-1-k-n + \frac{\omega}{2}, 1 - k -n + \frac{\omega}{2}, -2k-2n, -r^2 \right)  \right.
\nonumber\\
&+ \left. C_2 r^{2(k+n)} {}_2F_1 \left(k+n + \frac{\omega}{2}, 2+k+n+ \frac{\omega}{2}, 2(1+k+n), -r^2 \right) \right] \, .
\end{align}
Enforcing the boundary conditions proceeds in exactly the same manner as when studying the wave equation on AdS. To ensure regularity as $r \to 0$ we must set $C_2 = 0$. Then expanding near $r \to \infty$, the behaviour is
\begin{align} 
R(r) \sim&\, \frac{C_1  \Gamma \left[2(1+k + n) \right]}{\Gamma \left[2 + k + n - \omega/2 \right] \Gamma \left[2 + k + n + \omega/2 \right]} 
\nonumber\\
&+ \frac{C_1 \left[2(k+n) + \omega \right] \Gamma \left[2(1+k+n) \right]}{2 r^2 \Gamma\left[1+k+n-\omega/2\right]\Gamma\left[2+k+n+ \omega/2\right]} + \mathcal{O}(r^{-4}) \, .
\end{align}
To obtain the proper fall off, we must have the first two terms vanish, so that the radial solution decays at $\mathcal{O}(r^{-4})$. Obviously, we cannot set $C_1 = 0$, since this results in the trivial solution. Therefore, we use the property of the gamma function that $\Gamma[-N] = \infty$. Then, taking $\omega$ to be a positive quantity, we obtain the solution:
\be 
\omega^{(p=2)}_{N, k, n} = 2 \left(2 + k + n + N \right) \, .
\ee

Remarkably, this analytic result matches with numerical results, despite the fact that the limit $p \to 2$ is in a sense singular\footnote{Consider the behaviour in the vicinity of $r = a$. For $p > 2$, we have the behaviour
\be 
R(r) \sim (r-a)^{|n|/2}
\ee
while for $p = 2$, the behaviour is
\be 
R(r) \sim r^{2(k+|n|)} \, .
\ee
Obviously, there is a singular change in behaviour as $p \to 2$ (i.e. $a \to 0$), provided $n, k \neq 0$. }.  Numerical indications suggest that the eigenvalues $\omega$ of the radial equation have a well-defined limit as $p\to 2$ with result the same as that given just above:
\be\label{omegaAdS} 
\omega_{N, k, n}^{p\to 2^+} = \omega^{(p=2)}_{N, k, n} =   2\left(2 +k +n + N \right) \, ,
\ee
where $N$ is the overtone. We initially deduced this form by inspection, computing numerically values of $\omega$ as a function of $p$ in the close vicinity of $p=2$ for several values of $n$ and $k$.\footnote{For example, it is possible to obtain convergent results with the pseudospectral method for $p = 2 + 10^{-5}$ with a Chebychev discretization of about 500 points.} 

\begin{figure}[h]
\centering
\includegraphics[width=0.45\textwidth]{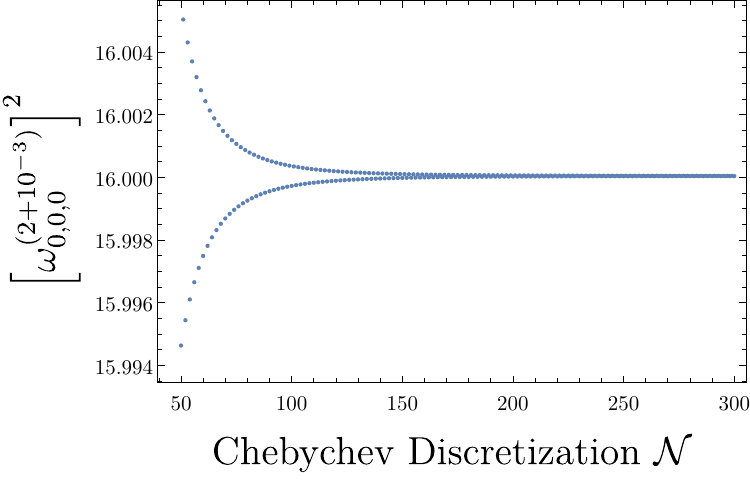}
\quad
\includegraphics[width=0.45\textwidth]{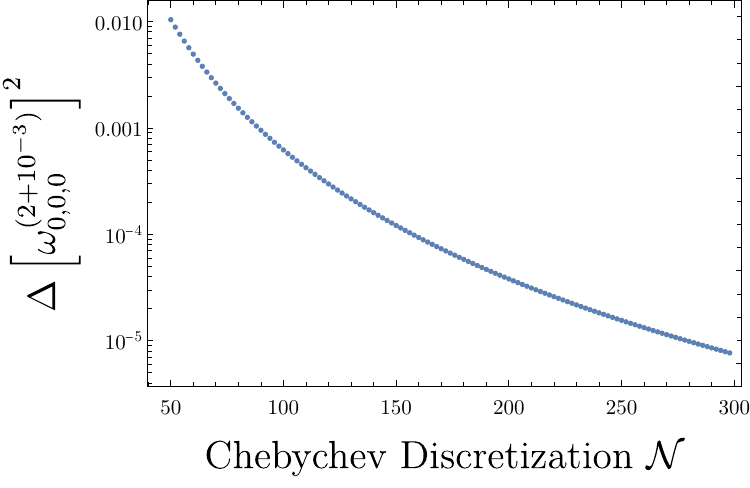}
\includegraphics[width=0.45\textwidth]{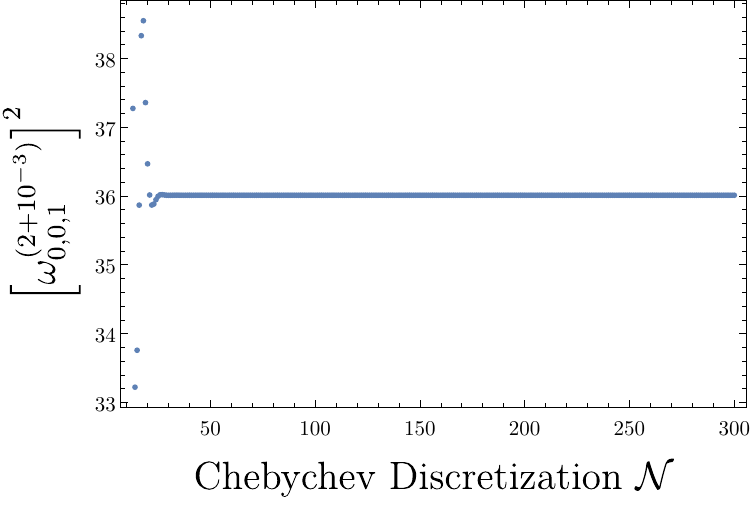}
\quad
\includegraphics[width=0.45\textwidth]{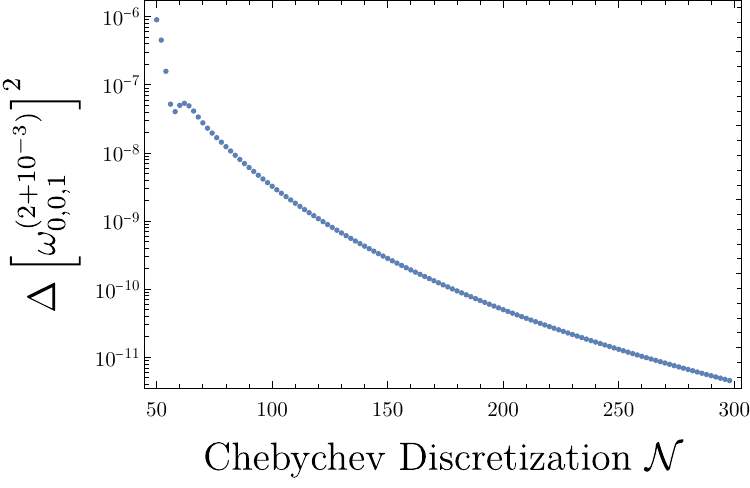}
\caption{Top left: A plot of the lowest eigenvalue for $n=0$, $k=0$ and $p=2+10^{-3}$ as a function of the number of points $\mathcal{N}$ used in the Chebychev discretization. The even and odd values of $N$ converge to the final result from opposite directions. Top right: A plot of the difference between the values of $\omega$ determined for even and odd $\mathcal{N}$ --- it is clear that the difference is quite rapidly approaching zero as the number of points used in the discretization is increased, indicating convergence. At $\mathcal{N} = 300$, the value is $\omega^2 = 16.000044$, with even and odd discretizations differing at order $10^{-6}$. The bottom row shows the same information, but now for the $n=1$ mode. In this case, convergence is more rapid. }
\label{p2converge}
\end{figure}

To highlight the convergence of the numerical scheme in the limit $p \to 2$, we include in Figure~\ref{p2converge} relevant plots. The top row depicts the case $n=0, k=0$ with $p = 2 +10^{-3}$, while in the bottom row the case $n=1, k=0$ is shown --- the results for different parameters are qualitatively similar. The top left plot shows the numerically computed value for the fundamental frequency as a function of the number of points used in the Chebychev discretization. The result converges to \eqref{omegaAdS} from above/below depending on whether the number of points is odd/even. In the top right plot, we show the difference between the numerical results computed with $\mathcal{N}$ and $\mathcal{N}+1$ points. The plot shows clearly the convergence. The plots included in the bottom row illustrate how, for $n \neq 0$, convergence is much more rapid (this has been a general feature of all our analysis here; it is not particular to the $p \to 2$ limit.)

\begin{figure}[h]
\centering
\includegraphics[width=0.45\textwidth]{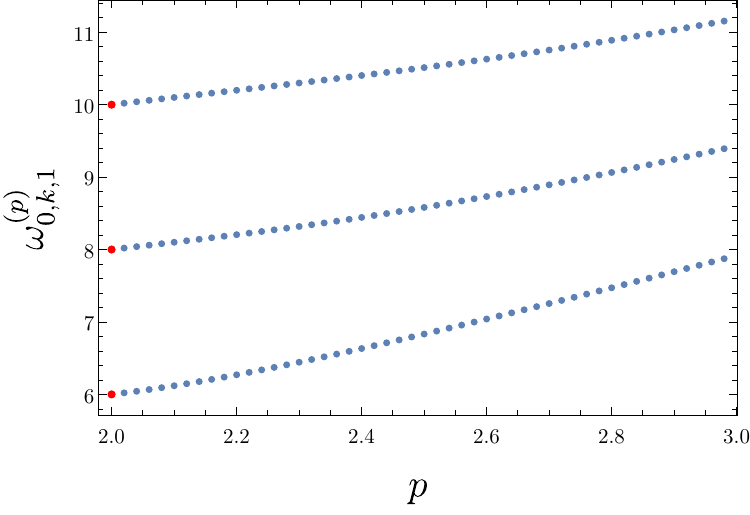}
\quad
\includegraphics[width=0.45\textwidth]{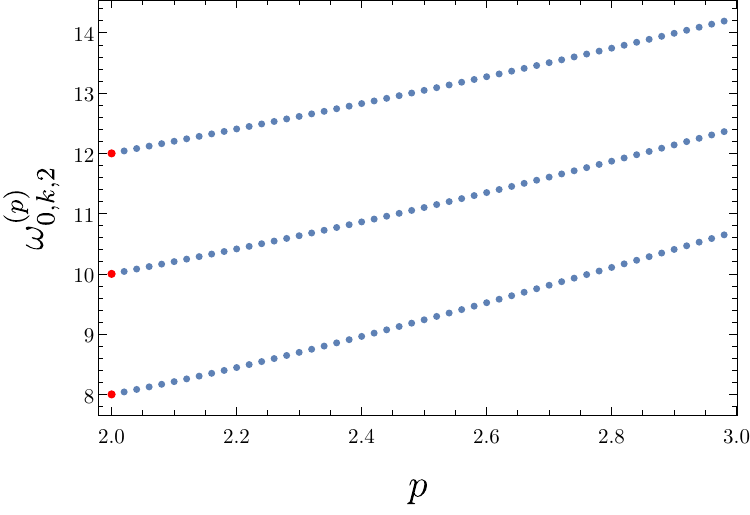}
\caption{Left: a plot of $\omega$ as a function of $p$ near $p=2$ for $n=1$ and $k=0,1,2$. The value of $k$ increases vertically between the different curves. Right: a plot of $\omega$ as a function of $p$ near $p=2$ for $n=2$ and $k=0,1,2$. The value of $k$ increases vertically between the different curves.  In all cases, the red dots at the far left indicate the value of $\omega$ at $p=2$ determined via analytic calculations.}
\label{p2approach}
\end{figure}

Finally, in Figure~\ref{p2approach}, we illustrate the dependence of the eigenvalues on $n$ and $k$ as a function of $p$ in the case when $p \to 2$. The left plot shows $n=1$ with the three curves corresponding to $k=0,1,2$ (bottom to top), while the right plot shows the same values of $k$ now for $n=2$. The structure of the curve is rather close to a linear dependence on $p$, but this is not as accurate as in the large $p$ limit.

\section{Tabulated Normal Mode Frequencies}
\label{modeTables}

Here we list the numerical values of the normal mode frequencies and the first eight overtones for $p = 3,4$ and $5$ for  $0 \le n \le 5$ and $0 \le k \le 5$.

\begin{table}[h]
\begin{center}
\begin{small}
\begin{tabular}{c c c c c c c c c c} 
 \toprule
 \multicolumn{10}{c}{Selected Normal Modes for Eguchi-Hanson Solitons with $p=3$} \\
 \toprule
 $n$ & $k$ & \multicolumn{8}{c}{$\ell \omega_{N, n, k} $} 
 \\
 \midrule
 0 & 0 & 5.29860 & 8.76887 & 12.26738 & 15.77064 & 19.27494 & 22.77952 & 26.28420 & 29.78891
 \\
 0 & 1 & 6.30468 & 9.36992 & 12.69502 & 16.10285 & 19.54661 & 23.00934 & 26.48335 & 29.96461
 \\
 0 & 2 & 7.92469 & 10.47381 & 13.51160 & 16.74837 & 20.07941 & 23.46252 & 26.87741 & 30.31310
 \\
 0 & 3 & 9.82910 & 11.95092 & 14.65595 & 17.67462 & 20.85428 & 24.12706 & 27.45838 & 30.82877
 \\
 0 & 4 & 11.84344 & 13.69088 & 16.06398 & 18.84279 & 21.84686 & 24.98695 & 28.21527 & 31.50380
 \\
 0 & 5 & 13.88455 & 15.60829 & 17.68099 & 20.21454 & 23.03088 & 26.02396 & 29.13512 & 32.32874
 \\ \midrule
 1 & 0 & 7.91778 & 11.14589 & 14.49899 & 17.90981 & 21.35104 & 24.81006 & 28.28034 & 31.75819
 \\
 1 & 1 & 9.42922 & 12.27359 & 15.38076 & 18.62934 & 21.95728 & 25.33320 & 28.74012 & 32.16812
 \\
 1 & 2 & 11.18450 & 13.69814 & 16.53884 & 19.59402 & 22.78002 & 26.04869 & 29.37222 & 32.73377
 \\
 1 & 3 & 13.06839 & 15.33579 & 17.92177 & 20.77130 & 23.79787 & 26.94192 & 30.16632 & 33.44760
 \\
 1 & 4 & 15.01696 & 17.12117 & 19.48357 & 22.12918 & 24.98833 & 27.99678 & 31.11066 & 34.30082
 \\
 1 & 5 & 16.99615 & 19.00469 & 21.18562 & 23.63863 & 26.32954 & 29.19684 & 32.19277 & 35.28387 
 \\ \midrule
 2 & 0 & 10.70812 & 13.74673 & 16.95599 & 20.26005 & 23.62075 & 27.01752 & 30.43849 & 33.87643
 \\
 2 & 1 & 12.41615 & 15.15809 & 18.13273 & 21.25980 & 24.48596 & 27.77830 & 31.11641 & 34.48725
 \\
 2 & 2 & 14.24590 & 16.75210 & 19.50481 & 22.44836 & 25.52750 & 28.70191 & 31.94433 & 35.23645
 \\
 2 & 3 & 16.14687 & 18.47772 & 21.03384 & 23.79816 & 26.72551 & 29.77380 & 32.91141 & 36.11579
 \\
 2 & 4 & 18.08915 & 20.29630 & 22.68719 & 25.28394 & 28.06072 & 30.97927 & 34.00632 & 37.11648
 \\
 2 & 5 & 20.05521 & 22.17902 & 24.43771 & 26.88344 & 29.51531 & 32.30412 & 35.21778 & 38.22946
 \\  \midrule
 3 & 0 & 13.57626 & 16.47128 & 19.55515 & 22.75613 & 26.03311 & 29.36154 & 32.72616 & 36.11706
 \\
 3 & 1 & 15.38154 & 18.04874 & 20.92806 & 23.95921 & 27.09791 & 30.31357 & 33.58533 & 36.89890
 \\
 3 & 2 & 17.25647 & 19.74496 & 22.44133 & 25.30787 & 28.30562 & 31.40240 & 34.57400 & 37.80273
 \\
 3 & 3 & 19.17545 & 21.52829 & 24.06727 & 26.78005 & 29.63908 & 32.61474 & 35.68177 & 38.82033
 \\
 3 & 4 & 21.12238 & 23.37482 & 25.78276 & 28.35616 & 31.08231 & 33.93772 & 36.89830 & 39.94336
 \\
 3 & 5 & 23.08710 & 25.26671 & 27.56875 & 30.01910 & 32.62084 & 35.35928 & 38.21360 & 41.16357
 \\ \midrule
 4 & 0 & 16.48583 & 19.27040 & 22.24843 & 25.35576 & 28.55222 & 31.81198 & 35.11796 & 38.45854
 \\
 4 & 1 & 18.34645 & 20.95265 & 23.75643 & 26.70864 & 29.77172 & 32.91802 & 36.12749 & 39.38560
 \\
 4 & 2 & 20.25075 & 22.71605 & 25.36763 & 28.17479 & 31.10726 & 34.13883 & 37.24843 & 40.41971
 \\
 4 & 3 & 22.18434 & 24.54022 & 27.06206 & 29.73690 & 32.54451 & 35.46273 & 38.47128 & 41.55312
 \\
 4 & 4 & 24.13770 & 26.40968 & 28.82319 & 31.37983 & 34.07039 & 36.87869 & 39.78682 & 42.77817
 \\
 4 & 5 & 26.10450 & 28.31272 & 30.63743 & 33.09049 & 35.67318 & 38.37654 & 41.18636 & 44.08744
 \\ \midrule
 5 & 0 & 19.41999 & 22.11767 & 25.00675 & 28.03092 & 31.15262 & 34.34625 & 37.59409 & 40.88356
 \\
 5 & 1 & 21.31516 & 23.87006 & 26.61105 & 29.49631 & 32.49325 & 35.57697 & 38.72861 & 41.93395
 \\
 5 & 2 & 23.23959 & 25.68024 & 28.29302 & 31.05090 & 33.92866 & 36.90416 & 39.95895 & 43.07815
 \\
 5 & 3 & 25.18453 & 27.53459 & 30.03810 & 32.68119 & 35.44702 & 38.31775 & 41.27663 & 44.30901
 \\
 5 & 4 & 27.14397 & 29.42265 & 31.83427 & 34.37546 & 37.03769 & 39.80841 & 42.67359 & 45.61964
 \\
 5 & 5 & 29.11376 & 31.33639 & 33.67164 & 36.12358 & 38.69119 & 41.36761 & 44.14231 & 47.00348
 \\
 \bottomrule
\end{tabular}
\end{small}
\end{center}
\label{modes_p3}
\end{table}

\begin{table}
\begin{small}
\begin{center}
\begin{tabular}{c c c c c c c c c c} 
 \toprule
 \multicolumn{10}{c}{Selected Normal Modes for Eguchi-Hanson Solitons with $p=4$} \\
 \toprule
 $n$ & $k$ & \multicolumn{8}{c}{$\ell \omega_{N, n, k} $} 
 \\
 \midrule
 0 & 0 & 6.93332 & 11.72096 & 16.48465 & 21.23594 & 25.98100 & 30.72257 & 35.46202 & 40.20008
 \\
 0 & 1 & 7.60124 & 12.11154 & 16.76178 & 21.45093 & 26.15667 & 30.87111 & 35.59069 & 40.31358
 \\
 0 & 2 &  8.78461 & 12.85744 & 17.30284 & 21.87462 & 26.50456 & 31.16608 & 35.84667 & 40.53964
 \\
 0 & 3 & 10.30489 & 13.90214 & 18.08440 & 22.49536 & 27.01807 & 31.60343 & 36.22728 & 40.87641
 \\
 0 & 4 & 12.03169 & 15.18497 & 19.07726 & 23.29755 & 27.68810 & 32.17741 & 36.72868 & 41.32121
 \\
 0 & 5 & 13.88412 & 16.65214 & 20.25081 & 24.26344 & 28.50374 & 32.88095 & 37.34608 & 41.87064
 \\
 \midrule
 1 & 0 & 10.24051 & 14.73250 & 19.33705 & 23.99139 & 28.67155 & 33.36672 & 38.07137 & 42.78236
 \\
 1 & 1 & 11.51463 & 15.63997 & 20.03509 & 24.55669 & 29.14584 & 33.77494 & 38.42953 & 43.10133
 \\
 1 & 2 & 13.01805 & 16.77399 & 20.92989 & 25.29093 & 29.76658 & 34.31176 & 38.90201 & 43.52302
 \\
 1 & 3 & 14.67832 & 18.09221 & 21.99768 & 26.18007 & 30.52492 & 34.97130 & 39.48473 & 44.04451
 \\
 1 & 4 & 16.44563 & 19.55760 & 23.21485 & 27.20908 & 31.41103 & 35.74686 & 40.17295 & 44.66234
 \\
 1 & 5 & 18.28645 & 21.13971 & 24.55950 & 28.36294 & 32.41454 & 36.63116 & 40.96143 & 45.37263
 \\
 \midrule
 2 & 0 & 13.83063 & 18.05755 & 22.48254 & 27.01100 & 31.59912 & 36.22446 & 40.87447 & 45.54163
 \\
 2 & 1 & 15.36757 & 19.27010 & 23.47076 & 27.84021 & 32.31142 & 36.84779 & 41.42809 & 46.03928
 \\
 2 & 2 & 17.02663 & 20.63114 & 24.60448 & 28.80398 & 33.14613 & 37.58225 & 42.08290 & 46.62951
 \\
 2 & 3 & 18.77399 & 22.11310 & 25.86469 & 29.88939 & 34.09436 & 38.42154 & 42.83431 & 47.30889
 \\
 2 & 4 & 20.58567 & 23.69313 & 27.23391 & 31.08382 & 35.14699 & 39.35901 & 43.67739 & 48.07369
 \\
 2 & 5 &22.44462 & 25.35262 & 28.69663 & 32.37531 & 36.29506 & 40.38791 & 44.60699 & 48.91994
 \\
 \midrule
 3 & 0 & 17.55790 & 21.56648 & 25.82210 & 30.22019 & 34.70659 & 39.25103 & 43.83548 & 48.44859
 \\
 3 & 1 & 19.23356 & 22.97093 & 27.01462 & 31.24942 & 35.60858 & 40.05209 & 44.55499 & 49.10106
 \\
 3 & 2 & 20.98243 & 24.47698 & 28.31610 & 32.38580 & 36.61238 & 40.94854 & 45.36342 & 49.83637
 \\
 3 & 3 & 22.78667 & 26.06675 & 29.71220 & 33.61852 & 37.70992 & 41.93430 & 46.25615 & 50.65094
 \\
 3 & 4 & 24.63314 & 27.72554 & 31.19021 & 34.93744 & 38.89332 & 43.00329 & 47.22844 & 51.54105
 \\
 3 & 5 & 26.51208 & 29.44134 & 32.73898 & 36.33320 & 40.15505 & 44.14953 & 48.27554 & 52.50291
 \\
 \midrule
 4 & 0 & 21.36191 & 25.19187 & 29.29583 & 33.56898 & 37.95281 & 42.41246 & 46.92611 & 51.47943
 \\
 4 & 1 & 23.12032 & 26.72433 & 30.63627 & 34.75182 & 39.00691 & 43.36073 & 47.78648 & 52.26597
 \\
 4 & 2 & 24.92639 & 28.32918 & 32.05982 & 36.02062 & 40.14580 & 44.39074 & 48.72473 & 53.12633
 \\
 4 & 3 & 26.76978 & 29.99453 & 33.55583 & 37.36661 & 41.36251 & 45.49695 & 49.73651 & 54.05703
 \\
 4 & 4 & 28.64264 & 31.71054 & 35.11495 & 38.78177 & 42.65040 & 46.67401 & 50.81744 & 55.05453
 \\
 4 & 5 & 30.53890 & 33.46910 & 36.72904 & 40.25880 & 44.00325 & 47.91671 & 51.96326 & 56.11529
 \\
 \midrule
 5 & 0 & 25.21339 & 28.89566 & 32.86589 & 37.02311 & 41.30781 & 45.68286 & 50.12401 & 54.61483
 \\
 5 & 1 & 27.02548 & 30.51774 & 34.31643 & 38.32580 & 42.48500 & 46.75368 & 51.10430 & 55.51758
 \\
 5 & 2 & 28.87050 & 32.19326 & 35.83160 & 39.69815 & 43.73313 & 47.89463 & 52.15280 & 56.48605
 \\
 5 & 3 & 30.74204 & 33.91404 & 37.40346 & 41.13315 & 45.04632 & 49.10086 & 53.26550 & 57.51693
 \\
 5 & 4 & 32.63507 & 35.67325 & 39.02504 & 42.62444 & 46.41905 & 50.36769 & 54.43849 & 58.60695
 \\
 5 & 5 & 34.54561 & 37.46519 & 40.69025 & 44.16628 & 47.84620 & 51.69068 & 55.66798 & 59.75290
 \\
 \bottomrule
\end{tabular}
\end{center}
\label{modes_p4}
\end{small}
\end{table}

\begin{table}
\begin{small}
\begin{center}
\begin{tabular}{c c c c c c c c c c} 
 \toprule
 \multicolumn{10}{c}{Selected Normal Modes for Eguchi-Hanson Solitons with $p=5$} \\
 \toprule
 $n$ & $k$ & \multicolumn{8}{c}{$\ell \omega_{N, n, k} $} 
 \\
 \midrule
 0 & 0 & 8.60470 & 14.66629 & 20.66753 & 26.64481 & 32.61081 & 38.57065 & 44.52678 & 50.48049
 \\
 0 & 1 & 9.11324 & 14.96233 & 20.87733 & 26.80748 & 32.74369 & 38.68298 & 44.62408 & 50.56631
 \\
 0 & 2 & 10.05321 & 15.53752 & 21.29074 & 27.12989 & 33.00784 & 38.90668 & 44.81805 & 50.73752
 \\
 0 & 3 & 11.31716 & 16.36255 & 21.89628 & 27.60647 & 33.40018 & 39.23985 & 45.10745 & 50.99326
 \\
 0 & 4 & 12.80896 & 17.40204 & 22.67861 & 28.22944 & 33.91625 & 39.67973 & 45.49046 & 51.33228
 \\
 0 & 5 & 14.45752 & 18.62026 & 23.62024 & 28.98940 & 34.55055 & 40.22285 & 45.96475 & 51.75293
 \\
 \midrule
 1 & 0 & 12.62102 & 18.34009 & 24.16444 & 30.03559 & 35.93107 & 41.84070 & 47.75927 & 53.68385
 \\
 1 & 1 & 13.78099 & 19.15265 & 24.78554 & 30.53704 & 36.35106 & 42.20180 & 48.07588 & 53.96567
 \\
 1 & 2 & 15.14170 & 20.14934 & 25.56199 & 31.16994 & 36.88404 & 42.66161 & 48.47991 & 54.32587
 \\
 1 & 3 & 16.65357 & 21.30438 & 26.48018 & 31.92650 & 37.52523 & 43.21697 & 48.96923 & 54.76291
 \\
 1 & 4 & 18.27861 & 22.59355 & 27.52596 & 32.79821 & 38.26919 & 43.86428 & 49.54130 & 55.27497
 \\
 1 & 5 & 19.98870 & 23.99529 & 28.68546 & 33.77618 & 39.11009 & 44.59954 & 50.19332 & 55.86000
 \\
 \midrule
 2 & 0 & 17.02736 & 22.41656 & 28.02660 & 33.75054 & 39.53958 & 45.36902 & 51.22514 & 57.09976
 \\
 2 & 1 & 18.48045 & 23.54323 & 28.93701 & 34.51085 & 40.19082 & 45.93785 & 51.72970 & 57.55290
 \\
 2 & 2 & 20.04220 & 24.79302 & 29.96427 & 35.37729 & 40.93753 & 46.59272 & 52.31222 & 58.07710
 \\
 2 & 3 & 21.68881 & 26.14828 & 31.09683 & 36.34228 & 41.77463 & 47.33009 & 52.97014 & 58.67048
 \\
 2 & 4 & 23.40202 & 27.59345 & 32.32364 & 37.39823 & 42.69681 & 48.14618 & 53.70070 & 59.33097
 \\
 2 & 5 & 25.16787 & 29.11511 & 33.63442 & 38.53767 & 43.69872 & 49.03708 & 54.50100 & 60.05635
 \\
 \midrule
 3 & 0 & 21.62601 & 26.73381 & 32.13452 & 37.70106 & 43.36904 & 49.10308 & 54.88238 & 60.69395
 \\
 3 & 1 & 23.23681 & 28.06247 & 33.25231 & 38.66034 & 44.20662 & 49.84506 & 55.54763 & 61.29643
 \\
 3 & 2 & 24.91336 & 29.47653 & 34.45875 & 39.70509 & 45.12438 & 50.66151 & 56.28186 & 61.96287
 \\
 3 & 3 & 26.64301 & 30.96426 & 35.74486 & 40.82877 & 46.11756 & 51.54890 & 57.08243 & 62.69125
 \\
 3 & 4 & 28.41581 & 32.51548 & 37.10236 & 42.02508 & 47.18140 & 52.50364 & 57.94659 & 63.47945
 \\
 3 & 5 & 30.22393 & 34.12150 & 38.52371 & 43.28799 & 48.31124 & 53.52215 & 58.87156 & 64.32526
 \\
 \midrule
 4 & 0 & 26.33349 & 31.20580 & 36.41467 & 41.82709 & 47.37057 & 53.00282 & 58.69778 & 64.43856
 \\
 4 & 1 & 28.03963 & 32.67149 & 37.68495 & 42.94117 & 48.35921 & 53.88953 & 59.50052 & 65.17118
 \\
 4 & 2 & 29.78887 & 34.19866 & 39.02355 & 44.12449 & 49.41520 & 54.84053 & 60.36408 & 65.96115
 \\
 4 & 3 & 31.57388 & 35.77939 & 40.42368 & 45.37162 & 50.53434 & 55.85255 & 61.28592 & 66.80645
 \\
 4 & 4 & 33.38873 & 37.40685 & 41.87916 & 46.67746 & 51.71253 & 56.92233 & 62.26344 & 67.70499
 \\
 4 & 5 & 35.22865 & 39.07512 & 43.38440 & 48.03724 & 52.94584 & 58.04671 & 63.29407 & 68.65470
 \\
 \midrule
 5 & 0 & 31.10896 & 35.78320 & 40.81996 & 46.08697 & 51.50826 & 57.03751 & 62.64499 & 68.31085
 \\
 5 & 1 & 32.87760 & 37.34615 & 42.20524 & 47.32327 & 52.62040 & 58.04577 & 63.56568 & 69.15707
 \\
 5 & 2 & 34.67621 & 38.95485 & 43.64412 & 48.61614 & 53.78933 & 59.10958 & 64.53997 & 70.05460
 \\
 5 & 3 & 36.50024 & 40.60382 & 45.13146 & 49.96119 & 55.01143 & 60.22602 & 65.56548 & 71.00152
 \\
 5 & 4 & 38.34592 & 42.28828 & 46.66260 & 51.35432 & 56.28326 & 61.39222 & 66.63984 & 71.99588
 \\
 5 & 5 & 40.21015 & 44.00411 & 48.23336 & 52.79171 & 57.60152 & 62.60539 & 67.76074 & 73.03575
 \\
 \bottomrule
\end{tabular}
\end{center}
\label{modes_p5}
\end{small}
\end{table}

\clearpage

\bibliographystyle{JHEP}
\bibliography{Gravities}

\end{document}